\documentclass[aps,pra,twocolumn, amsmath,amssymb,showpacs,superscriptaddress,citeautoscript,floatfix]{revtex4-2}

\usepackage{graphicx}
\usepackage{dcolumn}
\usepackage{bm}
\usepackage{graphicx}
\usepackage{enumitem}
\usepackage{appendix}
\usepackage{subfigure}

\begin{document}

\preprint{APS/123-QED}

\title{Security Loophole Induced by Photorefractive Effect in Continous-variable Quantum Key Distribution System}

\author{Zehao Zhou}
\affiliation{State Key Laboratory of Advanced Optical Communication Systems and Networks, Institute for Quantum Sensing and Information Processing, Shanghai Jiao Tong University, Shanghai 200240, China}

\author{Peng Huang}
\email{huang.peng@sjtu.edu.cn}
\affiliation{State Key Laboratory of Advanced Optical Communication Systems and Networks, Institute for Quantum Sensing and Information Processing, Shanghai Jiao Tong University, Shanghai 200240, China}
\affiliation{Shanghai Research Center for Quantum Sciences, Shanghai 201315, China}
\affiliation{Hefei National Laboratory, Hefei 230088, China}

\author{Tao Wang}
\affiliation{State Key Laboratory of Advanced Optical Communication Systems and Networks, Institute for Quantum Sensing and Information Processing, Shanghai Jiao Tong University, Shanghai 200240, China}
\affiliation{Shanghai Research Center for Quantum Sciences, Shanghai 201315, China}
\affiliation{Hefei National Laboratory, Hefei 230088, China}

\author{Guihua Zeng}
\email{ghzeng@sjtu.edu.cn}
\affiliation{State Key Laboratory of Advanced Optical Communication Systems and Networks, Institute for Quantum Sensing and Information Processing, Shanghai Jiao Tong University, Shanghai 200240, China}
\affiliation{Shanghai Research Center for Quantum Sciences, Shanghai 201315, China}
\affiliation{Hefei National Laboratory, Hefei 230088, China}
\affiliation{Shanghai XunTai Quantech Co., Ltd, Shanghai, 200241, China}

\date{\today}

\begin{abstract}
Modulators based on the Mach-Zehnder interferometer (MZI) structure are widely used in continuous-variable quantum key distribution (CVQKD) systems. MZI-based variable optical attenuator (VOA) and amplitude modulator can reshape the waveform and control the intensity of coherent state signal to realize secret key information modulation in CVQKD system. However, these devices are not ideal, internal and external effects like non-linear effect and temperature may degrade their performance. In this paper, we analyzed the security loophole of CVQKD under the photorefractive effect (PE), which originates from the crystal characteristic of lithium niobate (LN). It is found that the refractive index change of modulators because of PE may lead to an overestimate or underestimate of the final secret key rate. This allows Eve to perform further attacks like intercept-resend to get more secret key information. To solve this problem, several countermeasures are proposed, which can effectively eliminate potential risks.
\end{abstract}

\maketitle
 

\section{INTRODUCTION}

Quantum communication technologies is a fast-developing field that is continuously increasing in significance. Quantum key distribution (QKD) is one of the most widely researched techniques in quantum communication \cite{BB84, 1, 2, 3}. QKD systems modulate the key information on different types of quantum signals including single photons and the quadrature of quantum beam, respectively called discrete-variable QKD and continuous-variable QKD \cite{4, 5, 6, 7, 8, 9, 10}. Compared to discrete-variable QKD (DVQKD), continous-variable QKD (CVQKD) has apparent superiority in information transfer efficiency and implementation difficulty \cite{19, 20}. A large proportion of CVQKD research and implementation nowadays is based on Gaussian-modulated coherent states (GMCS) schemes. Through researchers' protracted and unremitting efforts, the theoretical security of GMCS-CVQKD under individual, collective, and coherent attacks has been proven \cite{8, 9, 10, GMCS-Hom-Re-Ind, GMCS-Hom-Fo-Ind, GMCS-Het-Re-Ind-2, GMCS-Het-Re-Ind-2, GMCS-Asymp-Gau-1, GMCS-Asymp-Gau-3, GMCS-Col}. GMCS-CVQKD has developed to maturation and is usually preferred because of its reliability. However, in practice, security in theory cannot guarantee the absolute security of the systems. An eavesdropper can exploit the imperfection of the actual devices to carry out side-channel attacks, and further steal (part of) the secret information \cite{review1, RevModPhys.81.1301}. Therefore, it is crucial to identify and solve the practical security problems. 

Currently, the proposed practical security problems of CVQKD have three orientations, respectively aiming at the transmitting end \cite{Attack-Transmitter-1, Attack-Transmitter-2, Attack-Transmitter-3, Attack-Transmitter-4}, local oscillator light \cite{Attack-LO-1, Attack-LO-2, Attack-LO-3}, and receiving end \cite{Attack-Receiver-1, Attack-Receiver-2, Attack-Receiver-3, Attack-Receiver-4}. All of these attacks have corresponding defense methods put forward as well. Particularly, researchers bring forward a Local LO (LLO) scheme \cite{16, 17} that generates LO at the receiving end directly. LLO scheme avoids the channel transmission process of LO and completely eliminates the security problems aiming at the transmitted local oscillator (LO) light. For the security problems aimed at the receiving end, researchers also established a measurement-device-independent (MDI) scheme that can defend all side-channel attacks against the detector \cite{CV-MDI-QKD-1, CV-MDI-QKD-3, Pirandola2015}. These innovative results remarkably contribute to the security of CVQKD which eradicates a series of loopholes. However, attacks with other targets like modulation devices are still threatening.

Mach-Zehnder interferometer is a basic optical structure commonly seen in optical devices \cite{MZM}. MZI modulators especially lithium niobate ($LiNbO_3$, LN) modulators are used in optical communication on a large scale. Lithium niobate crystals have many photoelectric effects, including piezoelectric effect, electrooptical effect, nonlinear optical effect, photovoltaic effect, photoelastic effect, photorefractive effect, and so on \cite{LN}. The excellent characteristics of the electrooptical effect and nonlinear optical effect make it favored in external modulators, like variable optical attenuator (VOA),  amplitude modulator (AM), and phase modulator (PM). Similarly, other features of LN may also influence the performance in both positive and negative way. The photorefractive effect is another one that extensively exists in LN materials \cite{PE2, PE4, PE5, Mechanism}. It describes the phenomenon of refractive index modulation of the waveguide under light irradiation. This index change can recover under specific conditions like heating. Although characteristics of the photorefractive effect may be useful in the field of data storage \cite{PE-Positive}, it may introduce adverse impacts on optical modulators as it's usually ignored. Recently, the security problems resulting from photorefractive effect are researched in DVQKD \cite{Attack-PE-1, Attack-PE-2}. However, the possible security loopholes under CVQKD have not been studied yet.

In this paper, the photorefractive effect in LN-based variable optical attenuators is studied. We introduce the generation mechanism and mathematical model of the photorefractive effect in CVQKD implementation. The negative impact of the photorefractive effect on modulators' transfer characteristics is explored. It is found that the transfer function of modulators deviates under different intensity of PE will lead to inaccurate signal output. Finally, security analysis and simulation results demonstrate that the secret key rate can be underestimated or overestimated by legal parties. This suggests that the photorefractive effect in amplitude modulators may create a security loophole for Eve to conceal her attacks. To eliminate the security risks caused by the photorefractive effect, we also proposed several effective countermeasures. However, in the long view, we further give some solutions to solving the problem in fundamental. In general, our work indicates an issue that was ignored regarding the devices' fundamental properties. Eavesdropper possibly gains key information and misleads the legal communication parties from this loophole. This discovery and corresponding proposed countermeasures contribute to perfecting the practical security of CVQKD systems, bringing the quantum communication techniques closer to absolute safety.

\section{PRINCIPLE OF THE LOOPHOLE FROM PHOTOREFRACTIVE EFFECT}

\subsection{Mechanism of photorefractive effect}

Compared with the instantaneous Kerr effect, the photorefractive effect is a time-continuing process. For electro-optical crystals. the impurities and defects can act as the donor or acceptor of electric charges. When these crystals are irradiated by uneven light, the charges (electron or hole) of impurities and defects are excited into the conduction or valence band to form charge carriers. These charge carriers then diffuse or drift due to the concentration gradient and external electric field applied. Besides, these charges may also move because of the photovoltaic effect \cite{PE4, PE5, PE2}. Therefore, the carriers are excited, migrated, and captured. The space charge distribution will be rearranged, and generate a space charge field. This space charge field modulates and changes the refractive index of the electro-optical crystal \cite{Attack-PE-1, Attack-PE-2, PE3, PE1}. 

For rigorous approach, the photorefractive process is described by a series of equations called Kukhtarev equation \cite{Kukhtarev}. According to the band transmission model, the generation rate of electrons under light excitation is $(N_D-N_D^+)(sI+\beta)$. Here $N_D$ is the donor density, $N_D^+$ is the ionized donor density, $s$ is the optical excitation constant, $I$ is the light intensity and $\beta$ is the thermal excitation probability. The recombination rate of electrons is $\gamma_RN_D^+\rho$, $\gamma_R$ is the recombination constant and $\rho$ is the electron density in the conduction band \cite{Mechanism}. The first Kukhtarev equation describing the change rate of charge carrier generation is
\begin{equation}
	\frac{\partial N_D^+}{\partial t}=(N_D-N_D^+)(sI+\beta)-\gamma_RN_D^+\rho,
\end{equation}
then the current density $J$ can be expressed by the second equation
\begin{equation}
	J=qD\bigtriangledown\rho+q\mu\rho E+pIe_c,
\end{equation}
where $D$ is the diffusion coefficient, $\mu$ is the electron mobility, $E$ is the electric field including external electric field and space-charge field, $p$ is the photovoltaic constant, $e_c$ is a direction vector. Three terms of the equation respectively represent diffusion current density, drift current density, and photovoltaic current density.  

The continuity equation is the third one describing the carrier current:
\begin{equation}
	\frac{\partial\rho}{\partial t}=\frac{\partial N_D^+}{\partial t}+\frac{\bigtriangledown\cdot J}{q}.
\end{equation}

The last modulation equation gives the relationship between refractive index change and space charge field $E_{sc}$ induced by PE:
\begin{equation}
	\bigtriangleup n(t)=-\frac{1}{2}n_0^3\gamma_{eff}r_{33}E_{sc}(t).
\end{equation} 
$n_0$ is the original refractive index, $\gamma_{eff}$ is the effective electro-optic coefficient, $r_{33}$ is the Pockels coefficient. As a time-continuing process, the photorefractive index change is a function of time $t$. Referring to Poisson's equation, the space charge field is
\begin{equation}
	E_{sc}(t)=(E_s-E_{sc}(0))(1-e^{-\frac{(\sigma_d+\sigma_{ph})t}{\epsilon\epsilon_0}})+E_{sc}(0).
\end{equation}

Assuming the space charge field induced by PE is zero at the beginning ($E_{sc}(0)=0$), the final space charge field will be $E_{sc}(\infty)=E_s$, where $E_s$ is the stable space charge field, or also called saturated space charge field, which is expressed by 
\begin{equation}
    E_{s} = \frac{\sigma_{ph}}{\sigma_d+\sigma_{ph}}E_{app}+\frac{\kappa\alpha}{\sigma_d+\sigma_{ph}}I_{ir}.
\label{E_s}
\end{equation}

Here $\sigma_d$ is the dark conductivity, $\sigma_{ph}$ is the photoconductivity, $\kappa$ is the Glass constant, and $\alpha$ is the absorption coefficient. 

Therefore, the stable or saturated refractive index change $\bigtriangleup n_s(\infty)$ is 
\begin{equation}
    \bigtriangleup n_{s}=-\frac{1}{2}n_0^3\gamma_{eff}r_{33}(\frac{\sigma_{ph}}{\sigma_d+\sigma_{ph}}E_{app}+\frac{\kappa\alpha}{\sigma_d+\sigma_{ph}}I_{ir})
\label{neff}.
\end{equation}

Then, the resulting phase deviation caused by refractive index change can be calculated with effective interaction length $L$, and signal wavelength $\lambda$ by
\begin{equation}
	\bigtriangleup\varphi_d=\frac{2\pi}{\lambda}\bigtriangleup n_{s}L.
\label{phid}
\end{equation}

It can be seen that the refractive index change of LN is jointly influenced by the intensity of uneven irradiation light and applied electric field. And this will further influence the quantum signal, leading to the abnormal output intensity of the modulator. 

\subsection{Impact on Mach-Zehnder modulator}

Variable optical attenuator is a type of Mach-Zehnder structure LN device that is widely used in optical communication systems. In the continuous-variable QKD system, VOA plays a similar role to an amplitude modulator that adjusts the output intensity of the signal by changing the relative phase between its two arms. Typically, the transfer characteristic of the Mach-Zehnder modulator (MZM) is presented in sinusoidal function. Take MXAN-LN series modulators from $\textit{iXblue Photonics}$ as an instance, the transfer function is expressed by 
\begin{equation}
    I_{out}=T_{mod}\cdot \frac{I_{in}}{2}[1+cos(\frac{\pi}{V_{\pi}}V_{DC}-\bigtriangleup\phi_0)],
\label{iout}
\end{equation}
where $T_{mod}$ is the optical transmittance of the device, $V_{\pi}$ is the half-wave voltage of the modulator, and $\phi_0$ is the phase deviation originates from the two arms of MZI because of fabrication error due to technological imperfections.

The analysis of the mechanical model indicates that the photorefractive effect on the Mach-Zehnder structure introduces a phase deviation $\bigtriangleup\varphi_d$ on each arm. We can define the overall phase deviation of the modulator as $\bigtriangleup\varphi_{p}=\bigtriangleup\varphi_{d(arm1)}-\bigtriangleup\varphi_{d(arm2)}$. Therefore, the transfer function of the modulator under the photorefractive effect is expressed by
\begin{equation}
    I'_{out}=T_{mod}\cdot \frac{I_{in}}{2}[1+cos(\frac{\pi}{V_{\pi}}V_{DC}-(\bigtriangleup\phi_0+\bigtriangleup\varphi_{p}))].
\label{i'out}
\end{equation}

For further analysis, define $k$ as an index called PE index to represent the intensity of photorefractive effect: 
\begin{equation}
	I'_{out}=kI_{out}.
\label{krelation}
\end{equation}

Since the phase deviation is derived from the applied electric field and irradiation intensity, the output intensity ratio $k$ can finally be calculated by input intensity $I_{in}$, applied voltage $V_{DC}$, and irradiation intensity $I_{ir}$. From Eq.(\ref{iout})-Eq.(\ref{krelation}), we can infer that
\begin{equation}
    k=\frac{I'_{out}}{I_{out}}=\frac{1+cos(\frac{\pi}{V_{\pi}}V_{DC}-(\bigtriangleup\phi_0+\bigtriangleup\varphi_{p}))}{1+cos(\frac{\pi}{V_{\pi}}V_{DC}-\bigtriangleup\phi_0)}.
\label{k}
\end{equation}

Substituting the Eq.(\ref{neff}) and Eq.(\ref{phid}), we know that 
\begin{equation}
    \bigtriangleup\varphi_d=\frac{e\mu\tau_0\eta_q\pi n_0^3 r_{33}\gamma I_{ir}(h\nu\kappa L-e\mu\tau_0\eta_q L_E E_{app})}{h^2 \nu^2 \lambda \sigma_d + h\nu e\mu\tau_0\eta_q\lambda\alpha I_{ir}},
\label{}
\end{equation}
where the photo-conductivity is linearly dependent on irradiation intensity as $\sigma_{ph}=e\mu\tau_0\eta_q\alpha I_{ir}/hv$. $e$ is the electronic charge, $\mu$ is the electron mobility, $\tau_0$ is the carrier lifetime, $\eta_q$ is the quantum efficiency, $h\nu$ is the photon energy, $L_E$ is the waveguide length modulated by the electric field.  

\begin{figure}
	\centering
	\subfigure[Transfer functions of Mach-Zehnder modulator under different applied voltage (zero at the full line)]{
		\begin{minipage}{0.4\textwidth} 
                        \includegraphics[width=\textwidth]{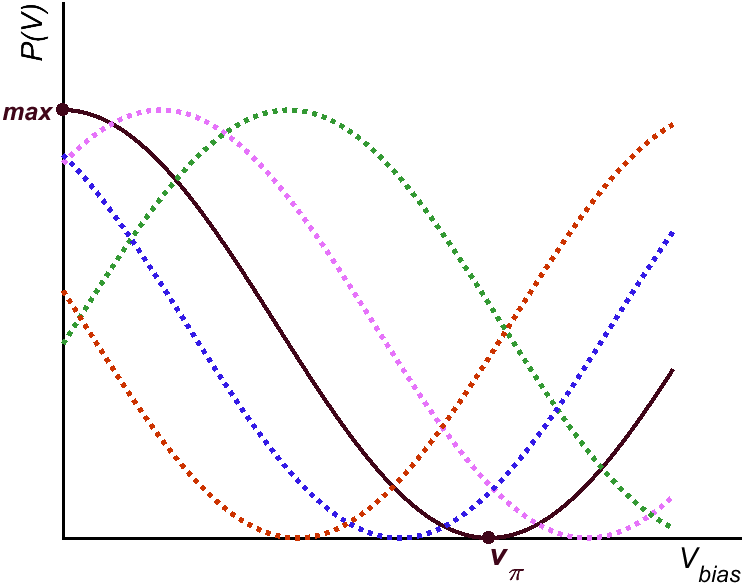} \\
		\end{minipage}
	}
	\subfigure[Transfer functions of modulator with and without photorefractive effect under a specific applied voltage]{
		\begin{minipage}{0.4\textwidth}
			\includegraphics[width=\textwidth]{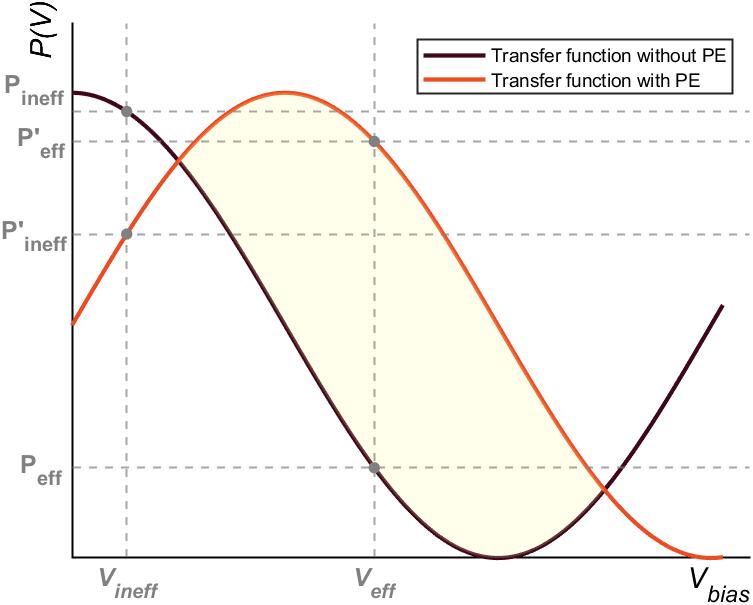} \\
		\end{minipage}
	}
 	\caption{Transfer function deviations of modulator} 
\label{test}
\end{figure}

Therefore, the overall phase deviation $\bigtriangleup\varphi_p$ of modulator under photorefractive effect can be calculated by taking the difference of phase deviation in two arms:
\begin{multline}
    \bigtriangleup\varphi_p=\frac{a \pi n^3 r_{33} \gamma}{\lambda_0 \sigma_d} \left[ L \left( \frac{I_{ir}^1}{1 + \frac{a \alpha}{\sigma_d} I_{ir}^1} - \frac{I_{ir}^2}{1 + \frac{a \alpha}{\sigma_d} I_{ir}^2} \right) \right. \\
        \left.  -\frac{a L_E}{d} \left( \frac{I_{ir}^1}{1 + \frac{a \alpha}{\sigma_d} I_{ir}^1} + \frac{I_{ir}^2}{1 + \frac{a \alpha}{\sigma_d} I_{ir}^2} \right) V_{app}  \right].
\label{phip}
\end{multline}

As Eq.(\ref{phip}) substituting into Eq.(\ref{k}), we can get the relationship between PE index $k$ and the controlled inputs: irradiation light intensity $I^{1,2}_{ir}$, and applied voltage $V_{app}$.

\begin{figure*}[]
	\centering
	\includegraphics[width=0.78\textwidth]{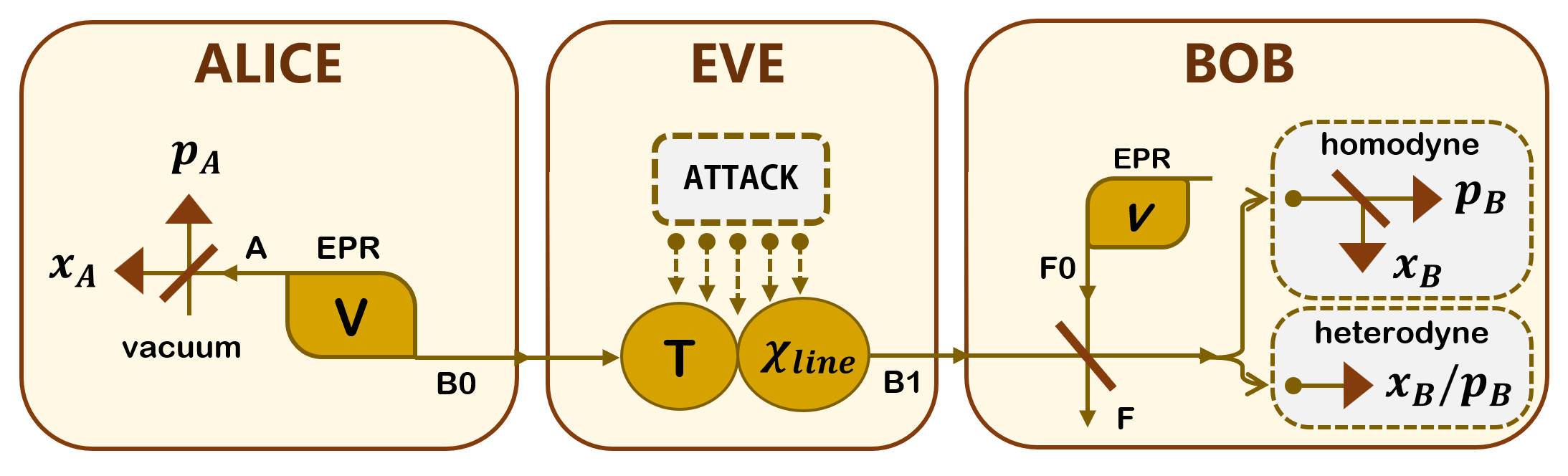}
	\caption{Entanglement-based model of GMCS-CVQKD protocol.}
        \label{EB model}
\end{figure*}

Research conducted in Ref.\cite{Attack-PE-1} revealed the characteristic change of VOA in discrete-variable QKD systems under the photorefractive effect. The photorefractive effect demonstrates identical consequences on MZM in CVQKD systems, as shown in Fig \ref{test}(a). The original transfer function of the modulator without PE is presented in full line. The power of the output signal reaches its maximum at zero bias voltage, and reaches its minimum at half-wave voltage. This conforms to Eq.(\ref{iout}), that the output signal intensity is the same as the input when the bias voltage is zero, as well as the output signal intensity is zero when the bias voltage is half-wave. The other dot line curves in different colors are the transfer functions under different irradiation light intensity and applied voltage. The output amplitude under PE has the same range but different at the same bias voltage. Thus, the impact of PE on MZM can be considered as the phase deviation in the transfer function.  

Fig \ref{test}(b) shows the comparison of two transfer function curves respectively with (red line) and without (black line) photorefractive effect. It is apparent that under certain operating voltage like $V_{eff}$, the attenuation effect of the modulator can be reduced, which means $k=P''_{eff}/P_{eff}>1$. And under some other operating voltages like $V_{ineff}$, the attenuation can also be strengthened, which means $k=P''_{ineff}/P_{ineff}<1$. This is the key feature of the photorefractive effect loophole reflecting on the system.
Different from the reduced optical attenuation effect proposed in Ref.\cite{Attack-Transmitter-3}, where $k$ is always larger than one due to the modulator thermal damage, the photorefractive effect index can be controlled either larger or less than one. Photorefractive effect is a reversible process though it always lasts long. This benefits Eve in hiding the loophole and provides more possibilities for attack patterns.

\section{PARAMETER ESTIMATION UNDER PHOTOREFRACTIVE EFFECT LOOPHOLE}

\subsection{Gaussian-modulated coherent state protocol}

One-way Gaussian-modulated coherent state CVQKD system is developed to relative maturation currently \cite{Implementation1, Implementation2, Huang2016, PhysRevLett.125.010502, Jouguet2013}. The working process of GMCS protocol is described by the prepare-and-measure model (PM model). Firstly, communication party Alice prepares a series of quantum coherent states with two orthogonal quadratures expressed by $x_i$ and $p_i$. They are modulated in Gaussian distribution with the same variance $V_A$ and zero mean value, and are sent to the other communication party Bob through the quantum channel. Then Bob measures one or both quadratures by homodyne detection or heterodyne detection. Due to modulation shot noise, channel excess noise and transfer efficiency, results are not completely the same as the initial states. Therefore, Bob and Alice will further share some information to perform the channel parameter estimation. The secure key rate of the system can be evaluated by estimated parameters. If the key rate is larger than zero, the communication system is secure in theory and the communication parties can finally perform data post-processing to get the final key. Otherwise, the communication will terminated and restarted.

In practice, the entanglement-based model (EB model) is always used to conduct security analysis. As shown in Fig \ref{EB model}, quantum coherent states prepared by Alice are equivalent to Einstein-Podolsky-Rosen (EPR) states. One mode $A$ is kept by Alice for heterodyne detection. The other mode $B_0$ is sent to Bob. The variance of the vacuum state here is $V=V_A+1$. At Bob's detection end, the practical efficiency of the detector is regarded as a beam splitter (BS) with transmittance $\eta$. And the electronic noise $e_vl$ is equivalent to the noise of an EPR state with variance $v$ going through BS. The variance is chosen to a value that does not change the total system noise.

\subsection{Process of parameter estimation}

In the GMCS-CVQKD protocol, Alice generates a series of Gaussian-modulated coherent states. It can be expressed as:
\begin{equation}
	|\alpha_A\rangle=|ae^{i\theta}\rangle=|x_A+ip_A\rangle.
\end{equation}

The quadrature $x_A=|a|cos\theta$ and $p_A=|a|sin\theta$. Both quadratures have a variance of $V_A$ and zero mean. It is obvious that $|a|^2$ is proportional to $I_A$, and $V_A=2\langle n \rangle$ proportional to $I_A$ as well. Therefore, the relationship between the quadratures and variance with and without photorefractive effect is
\begin{equation}
	x'_{Ao}=\sqrt{k}x_{Ao},
\end{equation}
\begin{equation}
	p'_{Ao}=\sqrt{k}p_{Ao},
\end{equation}
\begin{equation}
	V'_{Ao}=kV_{Ao}.
\end{equation}

In the GMCS-CVQKD protocol, the Gaussian attack has been proven to be optimal \cite{GMCS-Asymp-Gau-1, 8}. In this situation, the quantum signal channel is assumed to be linear, as described by 
\begin{equation}
    x_B=tx_{A_{o}}+z,
\end{equation}
where $t=\sqrt{\eta T}$ and $z$ is the normal distributed noise component with total variance $\sigma^2=\eta T\xi+N_0+V_{el}$. This implies the relations of the following parameters without photorefractive effect:
\begin{eqnarray}
	\left\{
	\begin{aligned}
		\langle x_{A_o}^2 \rangle &= V_{X_{A_o}} \\
		\langle x_{A_o} x_{B} \rangle &= \sqrt{\eta T} V_{X_{A_o}} \\
		\left\langle x_{B}^2 \right\rangle &= \eta T V_{X_{A_o}} + \eta T \xi + N_0 + V_{el},
	\end{aligned}
	\right.
\end{eqnarray}
$x_{A_o}$ and $x_{B}$ here represent the estimated parameters' value by Alice and Bob. Furthermore, the relationship still makes sense under a photorefractive attack. Therefore,
\begin{eqnarray}
	\left\{
	\begin{aligned}
		\langle {x^{'}_{A_o}}^2 \rangle &= V^{'}_{X_{A_o}} \\
		\langle x^{'}_{A_o} x^{'}_{B} \rangle &= \sqrt{\eta T} V^{'}_{X_{A_o}} \\
		\left\langle {x^{'}_{B}}^2 \right\rangle &= \eta T V^{'}_{X_{A_o}} + \eta T \xi + N_0 + V_{el}.
	\end{aligned}
	\right.
\end{eqnarray}

Here, $x^{'}_{A_o}$, $x^{'}_{B}$ and $V^{'}_{X_{A_o}}$ are practical values of parameters in the operating CVQKD system. However, for the evaluation of parameters, legal parties may not realize the attack and still use $x_{A_o}$ as they think, the relation becomes
\begin{eqnarray}
	\left\{
	\begin{aligned}
		\langle x_{A_o}^2 \rangle &= V_{X_{A_o}} \\
		\langle x_{A_o} x^{'}_{B} \rangle &= \sqrt{\eta T} V_{X_{A_o}} \\
		\left\langle {x^{'}_{B}}^2 \right\rangle &= \eta T V_{X_{A_o}} + \eta T \xi + N_0 + V_{el}.
	\end{aligned}
	\right.
\end{eqnarray}

Therefore, we have
\begin{equation}
    \sqrt{\mathbf{k}\eta T} \mathbf{V} x_{A_{0}} = \sqrt{\eta T'} \mathbf{V} x_{A_{0}},
\end{equation}
and
\begin{equation}
    \mathbf{k}\eta T \mathbf{V} x_{A_{0}} + \eta T \xi = \eta T' \mathbf{V} x_{A_{0}} + \eta T' \xi',
\end{equation}
which yields
\begin{eqnarray}
\begin{cases}
T' = k T \\
\xi' = \frac{\xi}{k} \quad \left( \varepsilon' = \frac{\varepsilon}{k} \right).
\end{cases}
\end{eqnarray}

Therefore, the practical transmittance of a one-way GMCS-CVQKD system under photorefractive effect is $k$ times of the transmittance without PE. While the excess noise with PE is one-$k^{th}$ of that without PE.

\section{SECURITY ANALYSIS AND KEY RATE UNDER PHOTOREFRACTIVE EFFECT}

The security analysis and secret key rate (SKR) calculation is implemented referring entanglement-based model as well. The security proof of one protocol is in fact proceeded by secret key rate calculation under specific attacks. Since collective attacks have been proven to be the most powerful attack in asymptotic conditions \cite{GMCS-Asymp-Gau-1, GMCS-Asymp-Gau-3, GMCS-Col}, a system can be considered as secure if the secure key rate is larger than zero under collective attacks. Specifically, in the case of reverse reconciliation, the secure key rate can be written as
\begin{equation}
R = f\cdot(\beta I_{AB} - \chi_{BE}),
\end{equation}
where $f$ is the repetition frequency of signal pulses, $\beta \in (0, 1)$ is the reverse reconciliation efficiency, $I_{AB}$ is the mutual information between Alice and Bob, and $\chi_{BE}$ is the maximum amount of information that Eve can extract from Bob's key. Without loss of generality, the following analysis of security mainly concentrates on homodyne detection.

For $I_{AB}$, under homodyne detection, it can be calculated by Shannon's equation with measurement variance and the conditional variance of Bob \cite{amplifier} as
\begin{equation}
I_{AB} = \frac{1}{2} \log_2 \frac{V_B}{V_{B|A}} = \frac{1}{2} \log_2 \frac{V + \chi_{\text{tot}}}{1 + \chi_{\text{tot}}},
\end{equation}    
where $V_B = \eta T (V + \chi_{\text{tot}})$ and $V_{B|A}=\eta T(1+\chi_{tot})$, $\chi_{tot}$ is the total noise referring to the channel input, it can be expressed as $\chi_{tot}=\chi_{line}+\chi_{hom}/T$. The total channel-added noise is further defined as $\chi_{line}=(1+T\epsilon)/T-1=1/T+\epsilon-1$. For the detection-added noise referring to Bob's input $\chi_{hom}$ can be calculated respectively as
\begin{equation}
    \chi_{hom}=[(1-\eta)+v_{el}]/\eta,
\end{equation}
where $\eta$ is the equivalent BS efficiency modeled by the EB model mentioned before, and $v_{el}$ is the electronic noise of the detector.

Then, estimating the upper bound of the information $\chi_{BE}$ that Eve can obtain is the core of secret key calculation. Under collective attacks, the Holevo bound is applied to estimate the maximum information Eve can extract \cite{GMCS-Asymp-Gau-1, GMCS-Asymp-Gau-3, Finitesize-1, Finitesize-2}, which can be acquired by
\begin{equation}
\chi_{BE} = \sum_{i=1}^{2} G \left( \frac{\lambda_i - 1}{2} \right) - \sum_{i=3}^{5} G \left( \frac{\lambda_i - 1}{2} \right),
\end{equation}
here $G(x) = (x+1) \log_2 (x+1) - x \log_2 x$ is the von Neumann entropy function.

The covariance matrix $\gamma_{AB_1}$ for the state $\rho_{AB_1}$ corresponds to the eigenvalues $\lambda_{1,2}$, while the covariance matrix $\gamma_{A B_1}$ for the state $\rho_{A B_1}^{m_B}$ corresponds to the eigenvalues $\lambda_{3,4,5}$.  The symplectic eigenvalues $\lambda_1$ and $\lambda_2$ can be calculated as
\begin{equation}
\lambda_{1,2} = \frac{1}{2} \left( A \pm \sqrt{A^2 - 4B} \right),
\end{equation}
where
\begin{align}
A &= V^2 (1 - 2T) + 2T + T^2 (V + \chi_{\text{line}})^2, \\
B &= T^2 (V \chi_{\text{line}} + 1)^2.
\end{align}

Similarly, analyzing the matrix $\gamma_{AFG}^{m_B}$, eigenvalues $\lambda_3$ and $\lambda_4$ can be calculated as follows:
$\gamma_{AFG}^{m_B}$. The eigenvalues are calculated as:
\begin{equation}
\lambda_{3,4} = \frac{1}{2} \left( C \pm \sqrt{C^2 - 4D} \right),
\end{equation}
where
\begin{align}
C_{\text{hom}} = \frac{A \chi_{\text{hom}} + V \sqrt{B} + T (V + \chi_{\text{line}})}{T (V + \chi_{\text{tot}})},\\
D_{\text{hom}} = \frac{\sqrt{B} V + \sqrt{B} \chi_{\text{hom}}}{T (V + \chi_{\text{tot}})}.
\end{align}

Therefore, the final secure key rate of the system can be calculated by modulation variance $V_A$, transmittance $T$, excess noise $\epsilon$, detection efficiency $\eta$, and electric noise $v_{el}$.  The final key rate is a function related to these parameters and so we can get a law that 
\begin{itemize}[label=\textbullet, font=\Large] 
    \item The practical security key rate without light injection attack $K$ is calculated by: $V_A, T, \epsilon, \eta, v_{el}$
    \item The practical security key rate with light injection attack $K_p$ is calculated by: $V'_A, T, \epsilon, \eta, v_{el}$
    \item The estimated security key rate with light injection attack $K_e$ evaluated by Alice and Bob is calculated by: $V_A, T', \epsilon ', \eta, v_{el}$
\end{itemize}

The brown and red surfaces are respectively the practical secret key rate and estimated secret key rate under a range of transmittance and PE index. Two surfaces intersect at the yellow dot line, where the PE index is 1. It is obvious that the practical SKR will be the same as the estimated SKR in this situation. When $k<1$, it can be seen that the practical SKR is always smaller than the estimated SKR. In this situation, the performance of the system is degraded. When $k>1$, the practical SKR is always larger than the estimated SKR. In this situation, SKR is overestimated, the estimated security key rate will be larger than the practical security key rate at the same distance. As well as Alice and Bob will underestimate the excess noise of the system, which means the estimated excess noise will be larger than the practical one. This creates a space for Eve allowing her to introduce additional excess noise to the system. Therefore, a security loophole is created that she can implement extra attacks like intercept-resend attacks to get the key information. Besides, it can be found that as $k$ becomes larger, the difference between practical SKR and estimated SKR gets larger too. This reflects that the leaking of the secret key information increases with a higher PE index. According to the above analysis, the optimal choice for Eve is to control the intensity of the photorefractive effect to keep the PE index larger than one ($k>1$). This also requires the bias voltage calibrated by users to be with the yellow region in Fig \ref{test}.

\begin{figure}[h]
	\centering
	\includegraphics[width=0.48\textwidth]{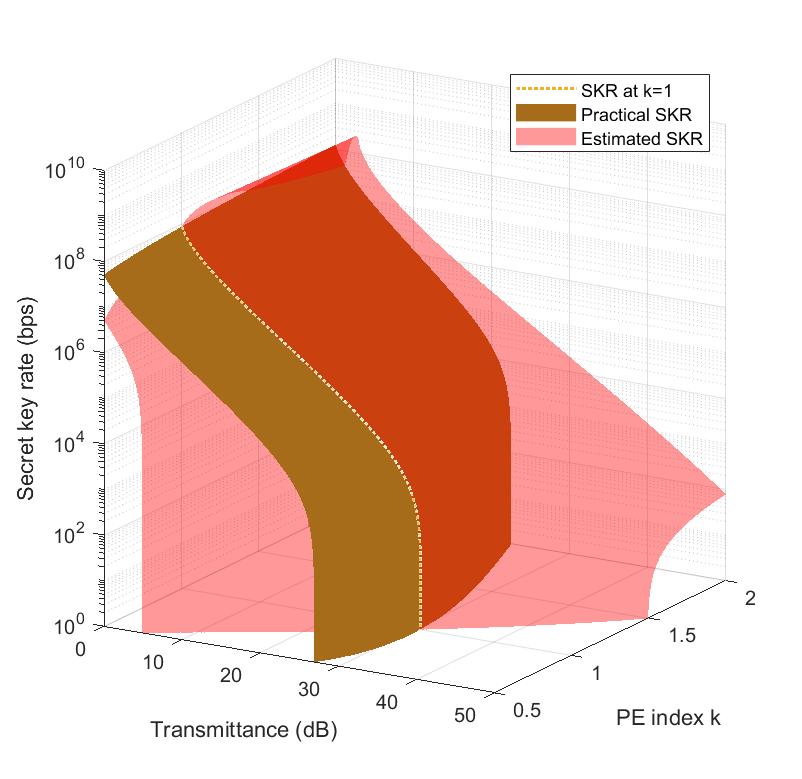}
	\caption{Practical and estimated secret key rate versus transmittance and PE index with 0.2 dB/km fiber loss, $f=300MHz$, $\eta=0.6$, $v_{el}=0.01$, $\varepsilon=0.05$, $V_A=4$, $\beta=0.95$.}
        \label{skrall}
\end{figure}

\begin{figure*}
	\centering
	\subfigure[Secret key rate at $k=0.8$. Curves from left to right respectively correspond to $\varepsilon=0.07, 0.05, 0.03.$]{
		\begin{minipage}{0.46\textwidth} 
                        \includegraphics[width=\textwidth]{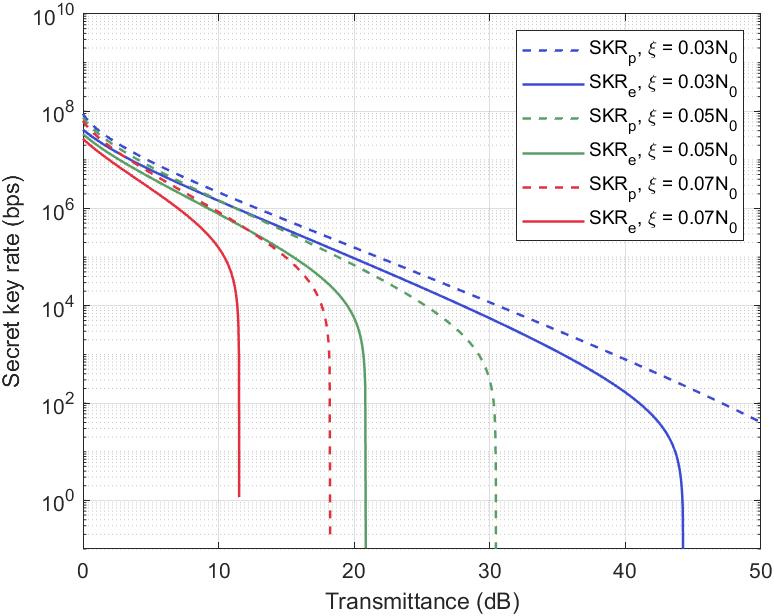} \\
		\end{minipage}
	}
	\subfigure[Secret key rate at $k=1.2$. Curves from left to right respectively correspond to $\varepsilon=0.07, 0.05, 0.03.$]{
		\begin{minipage}{0.46\textwidth}
			\includegraphics[width=\textwidth]{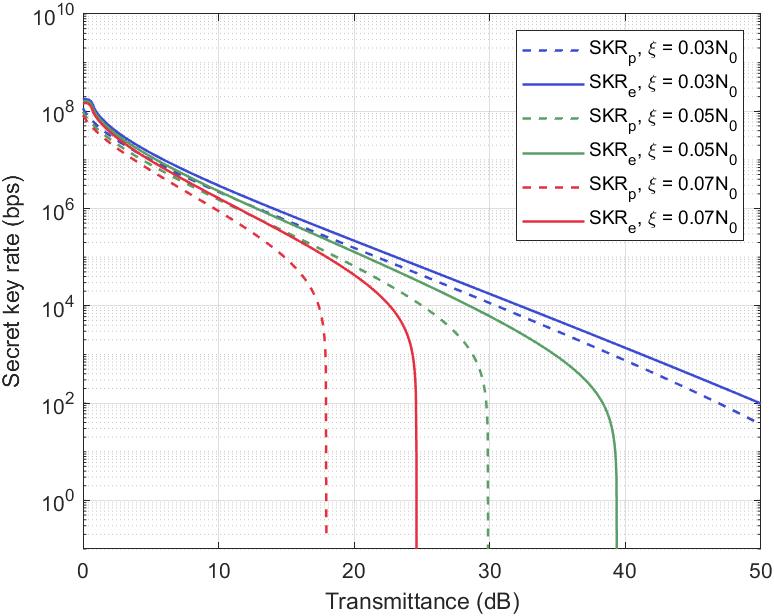} \\
		\end{minipage}
	}
 	\caption{Simulation results of secret key rate versus transmittance. Estimated result presented in solid line, practical result presented in dot line, parameters except excess noise keep unchanged.} 
  \label{skr}
\end{figure*}

Furthermore, Fig \ref{skr} shows the simulation results of secret key rate under different excess noise, respectively at $k=0.8$ and $k=1.2$. Both of the results indicate that the gap between practical and estimated SKR becomes bigger under the same PE index, as the excess noise becomes larger. This demonstrates that in the case of a larger excess noise, when $k<1$, Eve can degrade the communication performance to a greater extent; when $k>1$, Eve can acquire more secret key information.

\section{COUNTERMEASURES}

\subsection{Attack pattern}
Through the security analysis, it is found that the photorefractive effect that occurs on the modulator will open a security loophole. For Eve, there are different attack patterns to control the photorefractive effect of modulators.

\begin{figure}[h]
	\centering
	\includegraphics[width=0.48\textwidth]{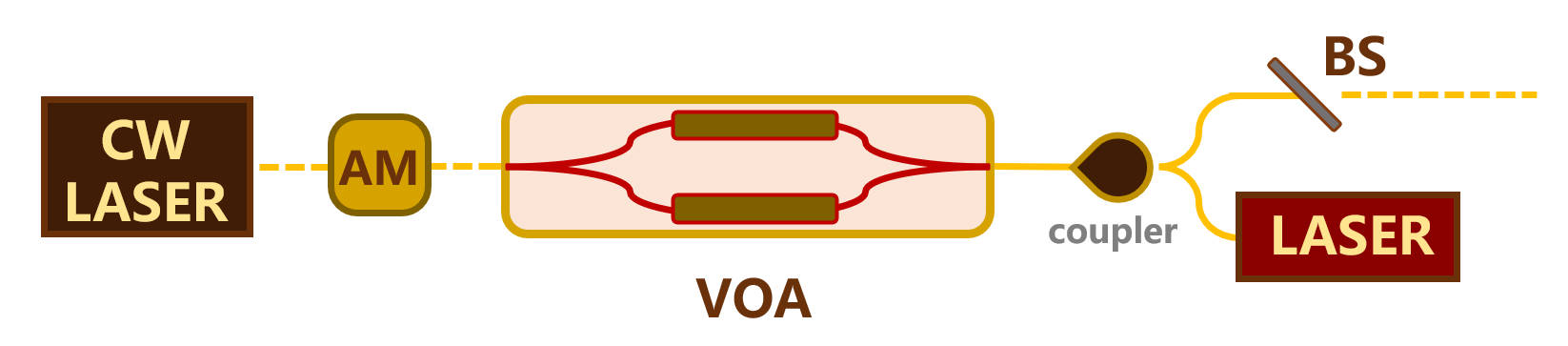}
	\caption{System diagram of traditional attack strategy}
        \label{attack}
\end{figure}

The first way to manipulate the modulator is to inject the light beam reversely into the modulators by coupling a laser. As shown in Fig \ref{attack}, the power of irradiation light can be adjusted by modulating the laser's pulse width. Under this pattern, the intensity of the photorefractive effect (PE index, $k$) is determined by irradiation power. Experiments in Ref.\cite{Attack-PE-1} have shown that the irradiation is even effective at only $3 nW$. For this reason, Eve can promise the modulator to be not damaged. 

The other way to create the loophole does not need to break the system. As mentioned, the photorefractive effect is a continuous process with a relatively long-term result. Generally, the photorefractive effect can last many weeks to several months in LN material \cite{PE1, PE2, PE3, PE4, PE5}. Therefore, Eve can act as the device supplier. She then processes the MZM under a specific electric field and irradiation to cause PE. This is done before providing it to Alice and Bob. The effect remains after the CVQKD system is established, and so that the loophole will exist originally in system. This way is more simple in implementation compared to the first one.

\subsection{Defense strategy}
Corresponding to the possible attack patterns, we also put forward some countermeasures. To fill the loophole, the direct solution is adding isolators or circulators after the modulator. However, the effect of isolation can still be weakened or destroyed under high reverse power. Besides, the real-time monitoring scheme can also have an effect. By monitoring the variation of the modulation variance in real-time, Alice and Bob can evaluate the PE index $k$ at any time. Therefore they can correct the parameters and get the proper secret key rate. Since the consequence of the PE loophole is similar to the reduced attenuation effect, the monitoring system can be designed referring to that in Ref.\cite{Attack-Transmitter-3} as well.

For the second pattern, the direct solution is to calibrate the modulators before putting them into service. However, this cannot fill the loophole perfectly as well. As time goes by, the recovery of the photorefractive effect will lead to the transfer function deviations again. The security loophole will reappear. Similar to the real-time monitoring system, it is effective to adjust the operating state of modulators in real-time. Therefore, users can promise the modulators are always in calibration. Currently, there is a mature product provided for automatic bias control (eg. iXblue MBC-DG-LAB). These controllers allow users to lock the operating point of LN Mach-Zehnder modulators, and to ensure a stable operation over time and environmental conditions \cite{biaslocking}. It is suggested to use the modulators jointly with this kind of automatic bias controller.

In long-term consideration, addressing the loophole in fundamental is preferred rather than just filling the loophole. The ultimate countermeasure is to eliminate the photorefractive effect. Optical cleaning is a way proposed in Ref.\cite{cleaning}. When the LN crystal is raised at approximately $180^\circ C$, the light beam is able to excite ions to transfer outside the illuminated region. Then these ions compensate the electronic
space-charge field and the illuminated region is cleared. This can significantly enhance the resistance to PE. 

Besides, the large amount of intrinsic defect due to the lack of lithium or non-stoichiometric structure of LN crystal is the fundamental reason for the photorefractive effect \cite{Doping3, Doping4}. Therefore, doping other elements in LN can also enhance the resistivity to PE. Until now, research has shown that the doping of magnesium, zinc, scandium, indium, hafnium, zirconium, and stannum can all inhibit the photorefractive effect \cite{Doping1, Doping2}. Especially $Mg$ doped LN and $Zr$ doped LN have high PE resistance and high crystalline quality. In view of this, the manufacturers are supposed to produce the modulators with doped LN devices in the future to completely close this PE security loophole. 

\section{CONCLUSION}
In this paper, we explore the photorefractive effect of LN crystals. Specifically, we studied the model of the effect and found its impact on LN-based Mach-Zehnder modulators. The transfer function of the amplitude modulator is deviated in phase. The intensity of PE $k$ is various and this related to the parameter estimation of the one-way GMCS-CVQKD system. The results of security analysis show that the secret key rate is overestimated by Alice and Bot, so that a security loophole is created when $k>1$. Eve is able to utilize the loophole to get the key information by intercept-resend attack. The leakage of key information is greater with larger system excess noise. Other than proposing the PE security loophole, we also put forward countermeasures against it. Communication parties can directly use isolators and calibrate modulators, or apply the variance real-time monitoring and automative bias controller. However, in the future, this security loophole may be closed by optical cleaning techniques and using doped-LN Mach-Zehnder modulators.  

\section{ACKNOWLEDGMENTS}
This work was supported by the Innovation Program for Quantum Science and Technology (Grant No. 2021ZD0300703), Shanghai Municipal Science and Technology Major Project (2019SHZDZX01), the Key R\&D Program of Guangdong province (Grant No. 2020B0303040002), and the National Natural Science Foundation of China (No. 62101320).
\nocite{*}

\bibliography{apssamp}

\providecommand{\noopsort}[1]{}\providecommand{\singleletter}[1]{#1}%
\begin{thebibliography}{70}%
\makeatletter
\providecommand \@ifxundefined [1]{%
 \@ifx{#1\undefined}
}%
\providecommand \@ifnum [1]{%
 \ifnum #1\expandafter \@firstoftwo
 \else \expandafter \@secondoftwo
 \fi
}%
\providecommand \@ifx [1]{%
 \ifx #1\expandafter \@firstoftwo
 \else \expandafter \@secondoftwo
 \fi
}%
\providecommand \natexlab [1]{#1}%
\providecommand \enquote  [1]{``#1''}%
\providecommand \bibnamefont  [1]{#1}%
\providecommand \bibfnamefont [1]{#1}%
\providecommand \citenamefont [1]{#1}%
\providecommand \href@noop [0]{\@secondoftwo}%
\providecommand \href [0]{\begingroup \@sanitize@url \@href}%
\providecommand \@href[1]{\@@startlink{#1}\@@href}%
\providecommand \@@href[1]{\endgroup#1\@@endlink}%
\providecommand \@sanitize@url [0]{\catcode `\\12\catcode `\$12\catcode
  `\&12\catcode `\#12\catcode `\^12\catcode `\_12\catcode `\%12\relax}%
\providecommand \@@startlink[1]{}%
\providecommand \@@endlink[0]{}%
\providecommand \url  [0]{\begingroup\@sanitize@url \@url }%
\providecommand \@url [1]{\endgroup\@href {#1}{\urlprefix }}%
\providecommand \urlprefix  [0]{URL }%
\providecommand \Eprint [0]{\href }%
\providecommand \doibase [0]{https://doi.org/}%
\providecommand \selectlanguage [0]{\@gobble}%
\providecommand \bibinfo  [0]{\@secondoftwo}%
\providecommand \bibfield  [0]{\@secondoftwo}%
\providecommand \translation [1]{[#1]}%
\providecommand \BibitemOpen [0]{}%
\providecommand \bibitemStop [0]{}%
\providecommand \bibitemNoStop [0]{.\EOS\space}%
\providecommand \EOS [0]{\spacefactor3000\relax}%
\providecommand \BibitemShut  [1]{\csname bibitem#1\endcsname}%
\let\auto@bib@innerbib\@empty
\bibitem [{\citenamefont {Bennett}\ and\ \citenamefont
  {Brassard}(2014)}]{BB84}%
  \BibitemOpen
  \bibfield  {author} {\bibinfo {author} {\bibfnamefont {C.~H.}\ \bibnamefont
  {Bennett}}\ and\ \bibinfo {author} {\bibfnamefont {G.}~\bibnamefont
  {Brassard}},\ }\bibfield  {title} {\bibinfo {title} {Quantum cryptography:
  Public key distribution and coin tossing},\ }\href
  {https://doi.org/https://doi.org/10.1016/j.tcs.2014.05.025} {\bibfield
  {journal} {\bibinfo  {journal} {Theoretical Computer Science}\ }\textbf
  {\bibinfo {volume} {560}},\ \bibinfo {pages} {7} (\bibinfo {year} {2014})},\
  \bibinfo {note} {theoretical Aspects of Quantum Cryptography – celebrating
  30 years of BB84}\BibitemShut {NoStop}%
\bibitem [{\citenamefont {Ekert}(1991)}]{1}%
  \BibitemOpen
  \bibfield  {author} {\bibinfo {author} {\bibfnamefont {A.~K.}\ \bibnamefont
  {Ekert}},\ }\bibfield  {title} {\bibinfo {title} {Quantum cryptography based
  on bell's theorem},\ }\href {https://doi.org/10.1103/PhysRevLett.67.661}
  {\bibfield  {journal} {\bibinfo  {journal} {Phys. Rev. Lett.}\ }\textbf
  {\bibinfo {volume} {67}},\ \bibinfo {pages} {661} (\bibinfo {year}
  {1991})}\BibitemShut {NoStop}%
\bibitem [{\citenamefont {Lo}\ and\ \citenamefont {Chau}(1998)}]{2}%
  \BibitemOpen
  \bibfield  {author} {\bibinfo {author} {\bibfnamefont {H.-K.}\ \bibnamefont
  {Lo}}\ and\ \bibinfo {author} {\bibfnamefont {H.~F.}\ \bibnamefont {Chau}},\
  }\bibfield  {title} {\bibinfo {title} {Unconditional security of quantum key
  distribution over arbitrarily long distances},\ }\href
  {https://api.semanticscholar.org/CorpusID:2948183} {\bibfield  {journal}
  {\bibinfo  {journal} {Science}\ }\textbf {\bibinfo {volume} {283 5410}},\
  \bibinfo {pages} {2050} (\bibinfo {year} {1998})}\BibitemShut {NoStop}%
\bibitem [{\citenamefont {Shor}\ and\ \citenamefont {Preskill}(2000)}]{3}%
  \BibitemOpen
  \bibfield  {author} {\bibinfo {author} {\bibfnamefont {P.~W.}\ \bibnamefont
  {Shor}}\ and\ \bibinfo {author} {\bibfnamefont {J.}~\bibnamefont
  {Preskill}},\ }\bibfield  {title} {\bibinfo {title} {Simple proof of security
  of the bb84 quantum key distribution protocol},\ }\href
  {https://doi.org/10.1103/PhysRevLett.85.441} {\bibfield  {journal} {\bibinfo
  {journal} {Phys. Rev. Lett.}\ }\textbf {\bibinfo {volume} {85}},\ \bibinfo
  {pages} {441} (\bibinfo {year} {2000})}\BibitemShut {NoStop}%
\bibitem [{\citenamefont {Ralph}(1999)}]{4}%
  \BibitemOpen
  \bibfield  {author} {\bibinfo {author} {\bibfnamefont {T.~C.}\ \bibnamefont
  {Ralph}},\ }\bibfield  {title} {\bibinfo {title} {Continuous variable quantum
  cryptography},\ }\href {https://doi.org/10.1103/PhysRevA.61.010303}
  {\bibfield  {journal} {\bibinfo  {journal} {Phys. Rev. A}\ }\textbf {\bibinfo
  {volume} {61}},\ \bibinfo {pages} {010303} (\bibinfo {year}
  {1999})}\BibitemShut {NoStop}%
\bibitem [{\citenamefont {Weedbrook}\ \emph {et~al.}(2004)\citenamefont
  {Weedbrook}, \citenamefont {Lance}, \citenamefont {Bowen}, \citenamefont
  {Symul}, \citenamefont {Ralph},\ and\ \citenamefont {Lam}}]{5}%
  \BibitemOpen
  \bibfield  {author} {\bibinfo {author} {\bibfnamefont {C.}~\bibnamefont
  {Weedbrook}}, \bibinfo {author} {\bibfnamefont {A.~M.}\ \bibnamefont
  {Lance}}, \bibinfo {author} {\bibfnamefont {W.~P.}\ \bibnamefont {Bowen}},
  \bibinfo {author} {\bibfnamefont {T.}~\bibnamefont {Symul}}, \bibinfo
  {author} {\bibfnamefont {T.~C.}\ \bibnamefont {Ralph}},\ and\ \bibinfo
  {author} {\bibfnamefont {P.~K.}\ \bibnamefont {Lam}},\ }\bibfield  {title}
  {\bibinfo {title} {Quantum cryptography without switching},\ }\href
  {https://doi.org/10.1103/PhysRevLett.93.170504} {\bibfield  {journal}
  {\bibinfo  {journal} {Phys. Rev. Lett.}\ }\textbf {\bibinfo {volume} {93}},\
  \bibinfo {pages} {170504} (\bibinfo {year} {2004})}\BibitemShut {NoStop}%
\bibitem [{\citenamefont {Hillery}(2000)}]{6}%
  \BibitemOpen
  \bibfield  {author} {\bibinfo {author} {\bibfnamefont {M.}~\bibnamefont
  {Hillery}},\ }\bibfield  {title} {\bibinfo {title} {Quantum cryptography with
  squeezed states},\ }\href {https://doi.org/10.1103/PhysRevA.61.022309}
  {\bibfield  {journal} {\bibinfo  {journal} {Phys. Rev. A}\ }\textbf {\bibinfo
  {volume} {61}},\ \bibinfo {pages} {022309} (\bibinfo {year}
  {2000})}\BibitemShut {NoStop}%
\bibitem [{\citenamefont {Bennett}\ \emph {et~al.}(1995)\citenamefont
  {Bennett}, \citenamefont {Brassard}, \citenamefont {Crepeau},\ and\
  \citenamefont {Maurer}}]{7}%
  \BibitemOpen
  \bibfield  {author} {\bibinfo {author} {\bibfnamefont {C.}~\bibnamefont
  {Bennett}}, \bibinfo {author} {\bibfnamefont {G.}~\bibnamefont {Brassard}},
  \bibinfo {author} {\bibfnamefont {C.}~\bibnamefont {Crepeau}},\ and\ \bibinfo
  {author} {\bibfnamefont {U.}~\bibnamefont {Maurer}},\ }\bibfield  {title}
  {\bibinfo {title} {Generalized privacy amplification},\ }\href
  {https://doi.org/10.1109/18.476316} {\bibfield  {journal} {\bibinfo
  {journal} {IEEE Transactions on Information Theory}\ }\textbf {\bibinfo
  {volume} {41}},\ \bibinfo {pages} {1915} (\bibinfo {year}
  {1995})}\BibitemShut {NoStop}%
\bibitem [{\citenamefont {Garc\'{\i}a-Patr\'on}\ and\ \citenamefont
  {Cerf}(2006)}]{8}%
  \BibitemOpen
  \bibfield  {author} {\bibinfo {author} {\bibfnamefont {R.}~\bibnamefont
  {Garc\'{\i}a-Patr\'on}}\ and\ \bibinfo {author} {\bibfnamefont {N.~J.}\
  \bibnamefont {Cerf}},\ }\bibfield  {title} {\bibinfo {title} {Unconditional
  optimality of gaussian attacks against continuous-variable quantum key
  distribution},\ }\href {https://doi.org/10.1103/PhysRevLett.97.190503}
  {\bibfield  {journal} {\bibinfo  {journal} {Phys. Rev. Lett.}\ }\textbf
  {\bibinfo {volume} {97}},\ \bibinfo {pages} {190503} (\bibinfo {year}
  {2006})}\BibitemShut {NoStop}%
\bibitem [{\citenamefont {Leverrier}(2015)}]{9}%
  \BibitemOpen
  \bibfield  {author} {\bibinfo {author} {\bibfnamefont {A.}~\bibnamefont
  {Leverrier}},\ }\bibfield  {title} {\bibinfo {title} {Composable security
  proof for continuous-variable quantum key distribution with coherent
  states},\ }\href {https://doi.org/10.1103/PhysRevLett.114.070501} {\bibfield
  {journal} {\bibinfo  {journal} {Phys. Rev. Lett.}\ }\textbf {\bibinfo
  {volume} {114}},\ \bibinfo {pages} {070501} (\bibinfo {year}
  {2015})}\BibitemShut {NoStop}%
\bibitem [{\citenamefont {Leverrier}(2017{\natexlab{a}})}]{10}%
  \BibitemOpen
  \bibfield  {author} {\bibinfo {author} {\bibfnamefont {A.}~\bibnamefont
  {Leverrier}},\ }\bibfield  {title} {\bibinfo {title} {Security of
  continuous-variable quantum key distribution via a gaussian de finetti
  reduction},\ }\bibfield  {journal} {\bibinfo  {journal} {Physical Review
  Letters}\ }\textbf {\bibinfo {volume} {118}},\ \href
  {https://doi.org/10.1103/physrevlett.118.200501}
  {10.1103/physrevlett.118.200501} (\bibinfo {year}
  {2017}{\natexlab{a}})\BibitemShut {NoStop}%
\bibitem [{\citenamefont {Pirandola}\ \emph {et~al.}(2017)\citenamefont
  {Pirandola}, \citenamefont {Laurenza}, \citenamefont {Ottaviani},\ and\
  \citenamefont {Banchi}}]{19}%
  \BibitemOpen
  \bibfield  {author} {\bibinfo {author} {\bibfnamefont {S.}~\bibnamefont
  {Pirandola}}, \bibinfo {author} {\bibfnamefont {R.}~\bibnamefont {Laurenza}},
  \bibinfo {author} {\bibfnamefont {C.}~\bibnamefont {Ottaviani}},\ and\
  \bibinfo {author} {\bibfnamefont {L.}~\bibnamefont {Banchi}},\ }\bibfield
  {title} {\bibinfo {title} {Fundamental limits of repeaterless quantum
  communications},\ }\bibfield  {journal} {\bibinfo  {journal} {Nature
  Communications}\ }\textbf {\bibinfo {volume} {8}},\ \href
  {https://doi.org/10.1038/ncomms15043} {10.1038/ncomms15043} (\bibinfo {year}
  {2017})\BibitemShut {NoStop}%
\bibitem [{\citenamefont {Zhang}\ \emph {et~al.}(2019)\citenamefont {Zhang},
  \citenamefont {Haw}, \citenamefont {Cai}, \citenamefont {Xu}, \citenamefont
  {Assad}, \citenamefont {Fitzsimons}, \citenamefont {Zhou}, \citenamefont
  {Zhang}, \citenamefont {Yu}, \citenamefont {Wu}, \citenamefont {Ser},
  \citenamefont {Kwek},\ and\ \citenamefont {Liu}}]{20}%
  \BibitemOpen
  \bibfield  {author} {\bibinfo {author} {\bibfnamefont {G.}~\bibnamefont
  {Zhang}}, \bibinfo {author} {\bibfnamefont {J.~Y.}\ \bibnamefont {Haw}},
  \bibinfo {author} {\bibfnamefont {H.}~\bibnamefont {Cai}}, \bibinfo {author}
  {\bibfnamefont {F.}~\bibnamefont {Xu}}, \bibinfo {author} {\bibfnamefont
  {S.~M.}\ \bibnamefont {Assad}}, \bibinfo {author} {\bibfnamefont {J.~F.}\
  \bibnamefont {Fitzsimons}}, \bibinfo {author} {\bibfnamefont
  {X.}~\bibnamefont {Zhou}}, \bibinfo {author} {\bibfnamefont {Y.}~\bibnamefont
  {Zhang}}, \bibinfo {author} {\bibfnamefont {S.}~\bibnamefont {Yu}}, \bibinfo
  {author} {\bibfnamefont {J.}~\bibnamefont {Wu}}, \bibinfo {author}
  {\bibfnamefont {W.}~\bibnamefont {Ser}}, \bibinfo {author} {\bibfnamefont
  {L.~C.}\ \bibnamefont {Kwek}},\ and\ \bibinfo {author} {\bibfnamefont
  {A.~Q.}\ \bibnamefont {Liu}},\ }\bibfield  {title} {\bibinfo {title} {An
  integrated silicon photonic chip platform for continuous-variable quantum key
  distribution},\ }\href {https://doi.org/10.1038/s41566-019-0504-5} {\bibfield
   {journal} {\bibinfo  {journal} {Nature Photonics}\ }\textbf {\bibinfo
  {volume} {13}},\ \bibinfo {pages} {839} (\bibinfo {year} {2019})}\BibitemShut
  {NoStop}%
\bibitem [{\citenamefont {Grosshans}\ and\ \citenamefont
  {Cerf}(2004)}]{GMCS-Hom-Re-Ind}%
  \BibitemOpen
  \bibfield  {author} {\bibinfo {author} {\bibfnamefont {F.}~\bibnamefont
  {Grosshans}}\ and\ \bibinfo {author} {\bibfnamefont {N.~J.}\ \bibnamefont
  {Cerf}},\ }\bibfield  {title} {\bibinfo {title} {Continuous-variable quantum
  cryptography is secure against non-gaussian attacks},\ }\bibfield  {journal}
  {\bibinfo  {journal} {Physical Review Letters}\ }\textbf {\bibinfo {volume}
  {92}},\ \href {https://doi.org/10.1103/physrevlett.92.047905}
  {10.1103/physrevlett.92.047905} (\bibinfo {year} {2004})\BibitemShut
  {NoStop}%
\bibitem [{\citenamefont {Grosshans}\ and\ \citenamefont
  {Grangier}(2002{\natexlab{a}})}]{GMCS-Hom-Fo-Ind}%
  \BibitemOpen
  \bibfield  {author} {\bibinfo {author} {\bibfnamefont {F.}~\bibnamefont
  {Grosshans}}\ and\ \bibinfo {author} {\bibfnamefont {P.}~\bibnamefont
  {Grangier}},\ }\bibfield  {title} {\bibinfo {title} {Continuous variable
  quantum cryptography using coherent states},\ }\href
  {https://doi.org/10.1103/PhysRevLett.88.057902} {\bibfield  {journal}
  {\bibinfo  {journal} {Phys. Rev. Lett.}\ }\textbf {\bibinfo {volume} {88}},\
  \bibinfo {pages} {057902} (\bibinfo {year} {2002}{\natexlab{a}})}\BibitemShut
  {NoStop}%
\bibitem [{\citenamefont {Lodewyck}\ and\ \citenamefont
  {Grangier}(2007)}]{GMCS-Het-Re-Ind-2}%
  \BibitemOpen
  \bibfield  {author} {\bibinfo {author} {\bibfnamefont {J.}~\bibnamefont
  {Lodewyck}}\ and\ \bibinfo {author} {\bibfnamefont {P.}~\bibnamefont
  {Grangier}},\ }\bibfield  {title} {\bibinfo {title} {Tight bound on the
  coherent-state quantum key distribution with heterodyne detection},\
  }\bibfield  {journal} {\bibinfo  {journal} {Physical Review A}\ }\textbf
  {\bibinfo {volume} {76}},\ \href {https://doi.org/10.1103/physreva.76.022332}
  {10.1103/physreva.76.022332} (\bibinfo {year} {2007})\BibitemShut {NoStop}%
\bibitem [{\citenamefont {Navascu\'es}\ \emph {et~al.}(2006)\citenamefont
  {Navascu\'es}, \citenamefont {Grosshans},\ and\ \citenamefont
  {Ac\'{\i}n}}]{GMCS-Asymp-Gau-1}%
  \BibitemOpen
  \bibfield  {author} {\bibinfo {author} {\bibfnamefont {M.}~\bibnamefont
  {Navascu\'es}}, \bibinfo {author} {\bibfnamefont {F.}~\bibnamefont
  {Grosshans}},\ and\ \bibinfo {author} {\bibfnamefont {A.}~\bibnamefont
  {Ac\'{\i}n}},\ }\bibfield  {title} {\bibinfo {title} {Optimality of gaussian
  attacks in continuous-variable quantum cryptography},\ }\href
  {https://doi.org/10.1103/PhysRevLett.97.190502} {\bibfield  {journal}
  {\bibinfo  {journal} {Phys. Rev. Lett.}\ }\textbf {\bibinfo {volume} {97}},\
  \bibinfo {pages} {190502} (\bibinfo {year} {2006})}\BibitemShut {NoStop}%
\bibitem [{\citenamefont {Wolf}\ \emph {et~al.}(2006)\citenamefont {Wolf},
  \citenamefont {Giedke},\ and\ \citenamefont {Cirac}}]{GMCS-Asymp-Gau-3}%
  \BibitemOpen
  \bibfield  {author} {\bibinfo {author} {\bibfnamefont {M.~M.}\ \bibnamefont
  {Wolf}}, \bibinfo {author} {\bibfnamefont {G.}~\bibnamefont {Giedke}},\ and\
  \bibinfo {author} {\bibfnamefont {J.~I.}\ \bibnamefont {Cirac}},\ }\bibfield
  {title} {\bibinfo {title} {Extremality of gaussian quantum states},\
  }\bibfield  {journal} {\bibinfo  {journal} {Physical Review Letters}\
  }\textbf {\bibinfo {volume} {96}},\ \href
  {https://doi.org/10.1103/physrevlett.96.080502}
  {10.1103/physrevlett.96.080502} (\bibinfo {year} {2006})\BibitemShut
  {NoStop}%
\bibitem [{\citenamefont {Grosshans}(2005)}]{GMCS-Col}%
  \BibitemOpen
  \bibfield  {author} {\bibinfo {author} {\bibfnamefont {F.}~\bibnamefont
  {Grosshans}},\ }\bibfield  {title} {\bibinfo {title} {Collectiveattacks and
  unconditional security in continuous variable quantum keydistribution},\
  }\bibfield  {journal} {\bibinfo  {journal} {Physical Review Letters}\
  }\textbf {\bibinfo {volume} {94}},\ \href
  {https://doi.org/10.1103/physrevlett.94.020504}
  {10.1103/physrevlett.94.020504} (\bibinfo {year} {2005})\BibitemShut
  {NoStop}%
\bibitem [{\citenamefont {Zhang}\ \emph {et~al.}(2024)\citenamefont {Zhang},
  \citenamefont {Bian}, \citenamefont {Li}, \citenamefont {Yu},\ and\
  \citenamefont {Guo}}]{review1}%
  \BibitemOpen
  \bibfield  {author} {\bibinfo {author} {\bibfnamefont {Y.}~\bibnamefont
  {Zhang}}, \bibinfo {author} {\bibfnamefont {Y.}~\bibnamefont {Bian}},
  \bibinfo {author} {\bibfnamefont {Z.}~\bibnamefont {Li}}, \bibinfo {author}
  {\bibfnamefont {S.}~\bibnamefont {Yu}},\ and\ \bibinfo {author}
  {\bibfnamefont {H.}~\bibnamefont {Guo}},\ }\bibfield  {title} {\bibinfo
  {title} {{Continuous-variable quantum key distribution system: Past, present,
  and future}},\ }\href {https://doi.org/10.1063/5.0179566} {\bibfield
  {journal} {\bibinfo  {journal} {Applied Physics Reviews}\ }\textbf {\bibinfo
  {volume} {11}},\ \bibinfo {pages} {011318} (\bibinfo {year} {2024})},\
  \Eprint
  {https://arxiv.org/abs/https://pubs.aip.org/aip/apr/article-pdf/doi/10.1063/5.0179566/19855574/011318\_1\_5.0179566.pdf}
  {https://pubs.aip.org/aip/apr/article-pdf/doi/10.1063/5.0179566/19855574/011318\_1\_5.0179566.pdf}
  \BibitemShut {NoStop}%
\bibitem [{\citenamefont {Scarani}\ \emph {et~al.}(2009)\citenamefont
  {Scarani}, \citenamefont {Bechmann-Pasquinucci}, \citenamefont {Cerf},
  \citenamefont {Du\ifmmode~\check{s}\else \v{s}\fi{}ek}, \citenamefont
  {L\"utkenhaus},\ and\ \citenamefont {Peev}}]{RevModPhys.81.1301}%
  \BibitemOpen
  \bibfield  {author} {\bibinfo {author} {\bibfnamefont {V.}~\bibnamefont
  {Scarani}}, \bibinfo {author} {\bibfnamefont {H.}~\bibnamefont
  {Bechmann-Pasquinucci}}, \bibinfo {author} {\bibfnamefont {N.~J.}\
  \bibnamefont {Cerf}}, \bibinfo {author} {\bibfnamefont {M.}~\bibnamefont
  {Du\ifmmode~\check{s}\else \v{s}\fi{}ek}}, \bibinfo {author} {\bibfnamefont
  {N.}~\bibnamefont {L\"utkenhaus}},\ and\ \bibinfo {author} {\bibfnamefont
  {M.}~\bibnamefont {Peev}},\ }\bibfield  {title} {\bibinfo {title} {The
  security of practical quantum key distribution},\ }\href
  {https://doi.org/10.1103/RevModPhys.81.1301} {\bibfield  {journal} {\bibinfo
  {journal} {Rev. Mod. Phys.}\ }\textbf {\bibinfo {volume} {81}},\ \bibinfo
  {pages} {1301} (\bibinfo {year} {2009})}\BibitemShut {NoStop}%
\bibitem [{\citenamefont {Huang}\ \emph
  {et~al.}(2013{\natexlab{a}})\citenamefont {Huang}, \citenamefont {He},\ and\
  \citenamefont {Zeng}}]{Attack-Transmitter-1}%
  \BibitemOpen
  \bibfield  {author} {\bibinfo {author} {\bibfnamefont {P.}~\bibnamefont
  {Huang}}, \bibinfo {author} {\bibfnamefont {G.}~\bibnamefont {He}},\ and\
  \bibinfo {author} {\bibfnamefont {G.}~\bibnamefont {Zeng}},\ }\bibfield
  {title} {\bibinfo {title} {Bound on noise of coherent source for secure
  continuous-variable quantum key distribution},\ }\href
  {https://doi.org/10.1007/s10773-012-1475-1} {\bibfield  {journal} {\bibinfo
  {journal} {International Journal of Theoretical Physics}\ }\textbf {\bibinfo
  {volume} {52}} (\bibinfo {year} {2013}{\natexlab{a}})}\BibitemShut {NoStop}%
\bibitem [{\citenamefont {Jain}\ \emph {et~al.}(2014)\citenamefont {Jain},
  \citenamefont {Anisimova}, \citenamefont {Khan}, \citenamefont {Makarov},
  \citenamefont {Marquardt},\ and\ \citenamefont
  {Leuchs}}]{Attack-Transmitter-2}%
  \BibitemOpen
  \bibfield  {author} {\bibinfo {author} {\bibfnamefont {N.}~\bibnamefont
  {Jain}}, \bibinfo {author} {\bibfnamefont {E.}~\bibnamefont {Anisimova}},
  \bibinfo {author} {\bibfnamefont {I.}~\bibnamefont {Khan}}, \bibinfo {author}
  {\bibfnamefont {V.}~\bibnamefont {Makarov}}, \bibinfo {author} {\bibfnamefont
  {C.}~\bibnamefont {Marquardt}},\ and\ \bibinfo {author} {\bibfnamefont
  {G.}~\bibnamefont {Leuchs}},\ }\bibfield  {title} {\bibinfo {title}
  {Trojan-horse attacks threaten the security of practical quantum
  cryptography},\ }\href {https://doi.org/10.1088/1367-2630/16/12/123030}
  {\bibfield  {journal} {\bibinfo  {journal} {New Journal of Physics}\ }\textbf
  {\bibinfo {volume} {16}},\ \bibinfo {pages} {123030} (\bibinfo {year}
  {2014})}\BibitemShut {NoStop}%
\bibitem [{\citenamefont {Zheng}\ \emph
  {et~al.}(2019{\natexlab{a}})\citenamefont {Zheng}, \citenamefont {Huang},
  \citenamefont {Huang}, \citenamefont {Peng},\ and\ \citenamefont
  {Zeng}}]{Attack-Transmitter-3}%
  \BibitemOpen
  \bibfield  {author} {\bibinfo {author} {\bibfnamefont {Y.}~\bibnamefont
  {Zheng}}, \bibinfo {author} {\bibfnamefont {P.}~\bibnamefont {Huang}},
  \bibinfo {author} {\bibfnamefont {A.}~\bibnamefont {Huang}}, \bibinfo
  {author} {\bibfnamefont {J.}~\bibnamefont {Peng}},\ and\ \bibinfo {author}
  {\bibfnamefont {G.}~\bibnamefont {Zeng}},\ }\bibfield  {title} {\bibinfo
  {title} {Practical security of continuous-variable quantum key distribution
  with reduced optical attenuation},\ }\href
  {https://doi.org/10.1103/PhysRevA.100.012313} {\bibfield  {journal} {\bibinfo
   {journal} {Phys. Rev. A}\ }\textbf {\bibinfo {volume} {100}},\ \bibinfo
  {pages} {012313} (\bibinfo {year} {2019}{\natexlab{a}})}\BibitemShut
  {NoStop}%
\bibitem [{\citenamefont {Zheng}\ \emph
  {et~al.}(2019{\natexlab{b}})\citenamefont {Zheng}, \citenamefont {Huang},
  \citenamefont {Huang}, \citenamefont {Peng},\ and\ \citenamefont
  {Zeng}}]{Attack-Transmitter-4}%
  \BibitemOpen
  \bibfield  {author} {\bibinfo {author} {\bibfnamefont {Y.}~\bibnamefont
  {Zheng}}, \bibinfo {author} {\bibfnamefont {P.}~\bibnamefont {Huang}},
  \bibinfo {author} {\bibfnamefont {A.}~\bibnamefont {Huang}}, \bibinfo
  {author} {\bibfnamefont {J.}~\bibnamefont {Peng}},\ and\ \bibinfo {author}
  {\bibfnamefont {G.}~\bibnamefont {Zeng}},\ }\bibfield  {title} {\bibinfo
  {title} {Security analysis of practical continuous-variable quantum key
  distribution systems under laser seeding attack},\ }\href
  {https://doi.org/10.1364/OE.27.027369} {\bibfield  {journal} {\bibinfo
  {journal} {Opt. Express}\ }\textbf {\bibinfo {volume} {27}},\ \bibinfo
  {pages} {27369} (\bibinfo {year} {2019}{\natexlab{b}})}\BibitemShut {NoStop}%
\bibitem [{\citenamefont {Ma}\ \emph {et~al.}(2013)\citenamefont {Ma},
  \citenamefont {Sun}, \citenamefont {Jiang},\ and\ \citenamefont
  {Liang}}]{Attack-LO-1}%
  \BibitemOpen
  \bibfield  {author} {\bibinfo {author} {\bibfnamefont {X.-C.}\ \bibnamefont
  {Ma}}, \bibinfo {author} {\bibfnamefont {S.-H.}\ \bibnamefont {Sun}},
  \bibinfo {author} {\bibfnamefont {M.-S.}\ \bibnamefont {Jiang}},\ and\
  \bibinfo {author} {\bibfnamefont {L.-M.}\ \bibnamefont {Liang}},\ }\bibfield
  {title} {\bibinfo {title} {Local oscillator fluctuation opens a loophole for
  eve in practical continuous-variable quantum-key-distribution systems},\
  }\href {https://doi.org/10.1103/PhysRevA.88.022339} {\bibfield  {journal}
  {\bibinfo  {journal} {Phys. Rev. A}\ }\textbf {\bibinfo {volume} {88}},\
  \bibinfo {pages} {022339} (\bibinfo {year} {2013})}\BibitemShut {NoStop}%
\bibitem [{\citenamefont {Jouguet}\ \emph
  {et~al.}(2013{\natexlab{a}})\citenamefont {Jouguet}, \citenamefont
  {Kunz-Jacques},\ and\ \citenamefont {Diamanti}}]{Attack-LO-2}%
  \BibitemOpen
  \bibfield  {author} {\bibinfo {author} {\bibfnamefont {P.}~\bibnamefont
  {Jouguet}}, \bibinfo {author} {\bibfnamefont {S.}~\bibnamefont
  {Kunz-Jacques}},\ and\ \bibinfo {author} {\bibfnamefont {E.}~\bibnamefont
  {Diamanti}},\ }\bibfield  {title} {\bibinfo {title} {Preventing calibration
  attacks on the local oscillator in continuous-variable quantum key
  distribution},\ }\href {https://doi.org/10.1103/PhysRevA.87.062313}
  {\bibfield  {journal} {\bibinfo  {journal} {Phys. Rev. A}\ }\textbf {\bibinfo
  {volume} {87}},\ \bibinfo {pages} {062313} (\bibinfo {year}
  {2013}{\natexlab{a}})}\BibitemShut {NoStop}%
\bibitem [{\citenamefont {Zhao}\ \emph {et~al.}(2018)\citenamefont {Zhao},
  \citenamefont {Zhang}, \citenamefont {Huang}, \citenamefont {Xu},
  \citenamefont {Yu},\ and\ \citenamefont {Guo}}]{Attack-LO-3}%
  \BibitemOpen
  \bibfield  {author} {\bibinfo {author} {\bibfnamefont {Y.}~\bibnamefont
  {Zhao}}, \bibinfo {author} {\bibfnamefont {Y.}~\bibnamefont {Zhang}},
  \bibinfo {author} {\bibfnamefont {Y.}~\bibnamefont {Huang}}, \bibinfo
  {author} {\bibfnamefont {B.}~\bibnamefont {Xu}}, \bibinfo {author}
  {\bibfnamefont {S.}~\bibnamefont {Yu}},\ and\ \bibinfo {author}
  {\bibfnamefont {H.}~\bibnamefont {Guo}},\ }\bibfield  {title} {\bibinfo
  {title} {Polarization attack on continuous-variable quantum key
  distribution},\ }\href {https://doi.org/10.1088/1361-6455/aaf0b7} {\bibfield
  {journal} {\bibinfo  {journal} {Journal of Physics B: Atomic, Molecular and
  Optical Physics}\ }\textbf {\bibinfo {volume} {52}},\ \bibinfo {pages}
  {015501} (\bibinfo {year} {2018})}\BibitemShut {NoStop}%
\bibitem [{\citenamefont {Qin}\ \emph {et~al.}(2018)\citenamefont {Qin},
  \citenamefont {Kumar}, \citenamefont {Makarov},\ and\ \citenamefont
  {All\'eaume}}]{Attack-Receiver-1}%
  \BibitemOpen
  \bibfield  {author} {\bibinfo {author} {\bibfnamefont {H.}~\bibnamefont
  {Qin}}, \bibinfo {author} {\bibfnamefont {R.}~\bibnamefont {Kumar}}, \bibinfo
  {author} {\bibfnamefont {V.}~\bibnamefont {Makarov}},\ and\ \bibinfo {author}
  {\bibfnamefont {R.}~\bibnamefont {All\'eaume}},\ }\bibfield  {title}
  {\bibinfo {title} {Homodyne-detector-blinding attack in continuous-variable
  quantum key distribution},\ }\href
  {https://doi.org/10.1103/PhysRevA.98.012312} {\bibfield  {journal} {\bibinfo
  {journal} {Phys. Rev. A}\ }\textbf {\bibinfo {volume} {98}},\ \bibinfo
  {pages} {012312} (\bibinfo {year} {2018})}\BibitemShut {NoStop}%
\bibitem [{\citenamefont {Wang}\ \emph {et~al.}(2016)\citenamefont {Wang},
  \citenamefont {Huang}, \citenamefont {Huang}, \citenamefont {Lin},\ and\
  \citenamefont {Zeng}}]{Attack-Receiver-2}%
  \BibitemOpen
  \bibfield  {author} {\bibinfo {author} {\bibfnamefont {C.}~\bibnamefont
  {Wang}}, \bibinfo {author} {\bibfnamefont {P.}~\bibnamefont {Huang}},
  \bibinfo {author} {\bibfnamefont {D.}~\bibnamefont {Huang}}, \bibinfo
  {author} {\bibfnamefont {D.}~\bibnamefont {Lin}},\ and\ \bibinfo {author}
  {\bibfnamefont {G.}~\bibnamefont {Zeng}},\ }\bibfield  {title} {\bibinfo
  {title} {Practical security of continuous-variable quantum key distribution
  with finite sampling bandwidth effects},\ }\href
  {https://doi.org/10.1103/PhysRevA.93.022315} {\bibfield  {journal} {\bibinfo
  {journal} {Phys. Rev. A}\ }\textbf {\bibinfo {volume} {93}},\ \bibinfo
  {pages} {022315} (\bibinfo {year} {2016})}\BibitemShut {NoStop}%
\bibitem [{\citenamefont {Qin}\ \emph {et~al.}(2016)\citenamefont {Qin},
  \citenamefont {Kumar},\ and\ \citenamefont {All\'eaume}}]{Attack-Receiver-3}%
  \BibitemOpen
  \bibfield  {author} {\bibinfo {author} {\bibfnamefont {H.}~\bibnamefont
  {Qin}}, \bibinfo {author} {\bibfnamefont {R.}~\bibnamefont {Kumar}},\ and\
  \bibinfo {author} {\bibfnamefont {R.}~\bibnamefont {All\'eaume}},\ }\bibfield
   {title} {\bibinfo {title} {Quantum hacking: Saturation attack on practical
  continuous-variable quantum key distribution},\ }\href
  {https://doi.org/10.1103/PhysRevA.94.012325} {\bibfield  {journal} {\bibinfo
  {journal} {Phys. Rev. A}\ }\textbf {\bibinfo {volume} {94}},\ \bibinfo
  {pages} {012325} (\bibinfo {year} {2016})}\BibitemShut {NoStop}%
\bibitem [{\citenamefont {Huang}\ \emph
  {et~al.}(2013{\natexlab{b}})\citenamefont {Huang}, \citenamefont {Weedbrook},
  \citenamefont {Yin}, \citenamefont {Wang}, \citenamefont {Li}, \citenamefont
  {Chen}, \citenamefont {Guo},\ and\ \citenamefont {Han}}]{Attack-Receiver-4}%
  \BibitemOpen
  \bibfield  {author} {\bibinfo {author} {\bibfnamefont {J.-Z.}\ \bibnamefont
  {Huang}}, \bibinfo {author} {\bibfnamefont {C.}~\bibnamefont {Weedbrook}},
  \bibinfo {author} {\bibfnamefont {Z.-Q.}\ \bibnamefont {Yin}}, \bibinfo
  {author} {\bibfnamefont {S.}~\bibnamefont {Wang}}, \bibinfo {author}
  {\bibfnamefont {H.-W.}\ \bibnamefont {Li}}, \bibinfo {author} {\bibfnamefont
  {W.}~\bibnamefont {Chen}}, \bibinfo {author} {\bibfnamefont {G.-C.}\
  \bibnamefont {Guo}},\ and\ \bibinfo {author} {\bibfnamefont {Z.-F.}\
  \bibnamefont {Han}},\ }\bibfield  {title} {\bibinfo {title} {Quantum hacking
  of a continuous-variable quantum-key-distribution system using a wavelength
  attack},\ }\href {https://doi.org/10.1103/PhysRevA.87.062329} {\bibfield
  {journal} {\bibinfo  {journal} {Phys. Rev. A}\ }\textbf {\bibinfo {volume}
  {87}},\ \bibinfo {pages} {062329} (\bibinfo {year}
  {2013}{\natexlab{b}})}\BibitemShut {NoStop}%
\bibitem [{\citenamefont {Qi}\ \emph {et~al.}(2015)\citenamefont {Qi},
  \citenamefont {Lougovski}, \citenamefont {Pooser}, \citenamefont {Grice},\
  and\ \citenamefont {Bobrek}}]{16}%
  \BibitemOpen
  \bibfield  {author} {\bibinfo {author} {\bibfnamefont {B.}~\bibnamefont
  {Qi}}, \bibinfo {author} {\bibfnamefont {P.}~\bibnamefont {Lougovski}},
  \bibinfo {author} {\bibfnamefont {R.}~\bibnamefont {Pooser}}, \bibinfo
  {author} {\bibfnamefont {W.}~\bibnamefont {Grice}},\ and\ \bibinfo {author}
  {\bibfnamefont {M.}~\bibnamefont {Bobrek}},\ }\bibfield  {title} {\bibinfo
  {title} {Generating the local oscillator ``locally'' in continuous-variable
  quantum key distribution based on coherent detection},\ }\href
  {https://doi.org/10.1103/PhysRevX.5.041009} {\bibfield  {journal} {\bibinfo
  {journal} {Phys. Rev. X}\ }\textbf {\bibinfo {volume} {5}},\ \bibinfo {pages}
  {041009} (\bibinfo {year} {2015})}\BibitemShut {NoStop}%
\bibitem [{\citenamefont {Soh}\ \emph {et~al.}(2015)\citenamefont {Soh},
  \citenamefont {Brif}, \citenamefont {Coles}, \citenamefont {L\"utkenhaus},
  \citenamefont {Camacho}, \citenamefont {Urayama},\ and\ \citenamefont
  {Sarovar}}]{17}%
  \BibitemOpen
  \bibfield  {author} {\bibinfo {author} {\bibfnamefont {D.~B.~S.}\
  \bibnamefont {Soh}}, \bibinfo {author} {\bibfnamefont {C.}~\bibnamefont
  {Brif}}, \bibinfo {author} {\bibfnamefont {P.~J.}\ \bibnamefont {Coles}},
  \bibinfo {author} {\bibfnamefont {N.}~\bibnamefont {L\"utkenhaus}}, \bibinfo
  {author} {\bibfnamefont {R.~M.}\ \bibnamefont {Camacho}}, \bibinfo {author}
  {\bibfnamefont {J.}~\bibnamefont {Urayama}},\ and\ \bibinfo {author}
  {\bibfnamefont {M.}~\bibnamefont {Sarovar}},\ }\bibfield  {title} {\bibinfo
  {title} {Self-referenced continuous-variable quantum key distribution
  protocol},\ }\href {https://doi.org/10.1103/PhysRevX.5.041010} {\bibfield
  {journal} {\bibinfo  {journal} {Phys. Rev. X}\ }\textbf {\bibinfo {volume}
  {5}},\ \bibinfo {pages} {041010} (\bibinfo {year} {2015})}\BibitemShut
  {NoStop}%
\bibitem [{\citenamefont {Li}\ \emph {et~al.}(2014)\citenamefont {Li},
  \citenamefont {Zhang}, \citenamefont {Xu}, \citenamefont {Peng},\ and\
  \citenamefont {Guo}}]{CV-MDI-QKD-1}%
  \BibitemOpen
  \bibfield  {author} {\bibinfo {author} {\bibfnamefont {Z.}~\bibnamefont
  {Li}}, \bibinfo {author} {\bibfnamefont {Y.-C.}\ \bibnamefont {Zhang}},
  \bibinfo {author} {\bibfnamefont {F.}~\bibnamefont {Xu}}, \bibinfo {author}
  {\bibfnamefont {X.}~\bibnamefont {Peng}},\ and\ \bibinfo {author}
  {\bibfnamefont {H.}~\bibnamefont {Guo}},\ }\bibfield  {title} {\bibinfo
  {title} {Continuous-variable measurement-device-independent quantum key
  distribution},\ }\href {https://doi.org/10.1103/PhysRevA.89.052301}
  {\bibfield  {journal} {\bibinfo  {journal} {Phys. Rev. A}\ }\textbf {\bibinfo
  {volume} {89}},\ \bibinfo {pages} {052301} (\bibinfo {year}
  {2014})}\BibitemShut {NoStop}%
\bibitem [{\citenamefont {Tian}\ \emph {et~al.}(2022)\citenamefont {Tian},
  \citenamefont {Wang}, \citenamefont {Liu}, \citenamefont {Du}, \citenamefont
  {Liu}, \citenamefont {Lu}, \citenamefont {Wang},\ and\ \citenamefont
  {Li}}]{CV-MDI-QKD-3}%
  \BibitemOpen
  \bibfield  {author} {\bibinfo {author} {\bibfnamefont {Y.}~\bibnamefont
  {Tian}}, \bibinfo {author} {\bibfnamefont {P.}~\bibnamefont {Wang}}, \bibinfo
  {author} {\bibfnamefont {J.}~\bibnamefont {Liu}}, \bibinfo {author}
  {\bibfnamefont {S.}~\bibnamefont {Du}}, \bibinfo {author} {\bibfnamefont
  {W.}~\bibnamefont {Liu}}, \bibinfo {author} {\bibfnamefont {Z.}~\bibnamefont
  {Lu}}, \bibinfo {author} {\bibfnamefont {X.}~\bibnamefont {Wang}},\ and\
  \bibinfo {author} {\bibfnamefont {Y.}~\bibnamefont {Li}},\ }\bibfield
  {title} {\bibinfo {title} {Experimental demonstration of continuous-variable
  measurement-device-independent quantum key distribution over optical fiber},\
  }\href {https://doi.org/10.1364/OPTICA.450573} {\bibfield  {journal}
  {\bibinfo  {journal} {Optica}\ }\textbf {\bibinfo {volume} {9}},\ \bibinfo
  {pages} {492} (\bibinfo {year} {2022})}\BibitemShut {NoStop}%
\bibitem [{\citenamefont {Pirandola}\ \emph {et~al.}(2015)\citenamefont
  {Pirandola}, \citenamefont {Ottaviani}, \citenamefont {Spedalieri},
  \citenamefont {Weedbrook}, \citenamefont {Braunstein}, \citenamefont {Lloyd},
  \citenamefont {Gehring}, \citenamefont {Jacobsen},\ and\ \citenamefont
  {Andersen}}]{Pirandola2015}%
  \BibitemOpen
  \bibfield  {author} {\bibinfo {author} {\bibfnamefont {S.}~\bibnamefont
  {Pirandola}}, \bibinfo {author} {\bibfnamefont {C.}~\bibnamefont
  {Ottaviani}}, \bibinfo {author} {\bibfnamefont {G.}~\bibnamefont
  {Spedalieri}}, \bibinfo {author} {\bibfnamefont {C.}~\bibnamefont
  {Weedbrook}}, \bibinfo {author} {\bibfnamefont {S.~L.}\ \bibnamefont
  {Braunstein}}, \bibinfo {author} {\bibfnamefont {S.}~\bibnamefont {Lloyd}},
  \bibinfo {author} {\bibfnamefont {T.}~\bibnamefont {Gehring}}, \bibinfo
  {author} {\bibfnamefont {C.~S.}\ \bibnamefont {Jacobsen}},\ and\ \bibinfo
  {author} {\bibfnamefont {U.~L.}\ \bibnamefont {Andersen}},\ }\bibfield
  {title} {\bibinfo {title} {High-rate measurement-device-independent quantum
  cryptography},\ }\href {https://doi.org/10.1038/nphoton.2015.83} {\bibfield
  {journal} {\bibinfo  {journal} {Nature Photonics}\ }\textbf {\bibinfo
  {volume} {9}},\ \bibinfo {pages} {397} (\bibinfo {year} {2015})}\BibitemShut
  {NoStop}%
\bibitem [{\citenamefont {Fu}\ \emph {et~al.}(2013{\natexlab{a}})\citenamefont
  {Fu}, \citenamefont {Zhang}, \citenamefont {Hraimel}, \citenamefont {Liu},\
  and\ \citenamefont {Shen}}]{MZM}%
  \BibitemOpen
  \bibfield  {author} {\bibinfo {author} {\bibfnamefont {Y.}~\bibnamefont
  {Fu}}, \bibinfo {author} {\bibfnamefont {X.}~\bibnamefont {Zhang}}, \bibinfo
  {author} {\bibfnamefont {B.}~\bibnamefont {Hraimel}}, \bibinfo {author}
  {\bibfnamefont {T.}~\bibnamefont {Liu}},\ and\ \bibinfo {author}
  {\bibfnamefont {D.}~\bibnamefont {Shen}},\ }\bibfield  {title} {\bibinfo
  {title} {Mach-zehnder: A review of bias control techniques for mach-zehnder
  modulators in photonic analog links},\ }\href
  {https://doi.org/10.1109/MMM.2013.2280332} {\bibfield  {journal} {\bibinfo
  {journal} {IEEE Microwave Magazine}\ }\textbf {\bibinfo {volume} {14}},\
  \bibinfo {pages} {102} (\bibinfo {year} {2013}{\natexlab{a}})}\BibitemShut
  {NoStop}%
\bibitem [{\citenamefont {Boes}\ \emph {et~al.}(2023)\citenamefont {Boes},
  \citenamefont {Chang}, \citenamefont {Langrock}, \citenamefont {Yu},
  \citenamefont {Zhang}, \citenamefont {Lin}, \citenamefont {Lončar},
  \citenamefont {Fejer}, \citenamefont {Bowers},\ and\ \citenamefont
  {Mitchell}}]{LN}%
  \BibitemOpen
  \bibfield  {author} {\bibinfo {author} {\bibfnamefont {A.}~\bibnamefont
  {Boes}}, \bibinfo {author} {\bibfnamefont {L.}~\bibnamefont {Chang}},
  \bibinfo {author} {\bibfnamefont {C.}~\bibnamefont {Langrock}}, \bibinfo
  {author} {\bibfnamefont {M.}~\bibnamefont {Yu}}, \bibinfo {author}
  {\bibfnamefont {M.}~\bibnamefont {Zhang}}, \bibinfo {author} {\bibfnamefont
  {Q.}~\bibnamefont {Lin}}, \bibinfo {author} {\bibfnamefont {M.}~\bibnamefont
  {Lončar}}, \bibinfo {author} {\bibfnamefont {M.}~\bibnamefont {Fejer}},
  \bibinfo {author} {\bibfnamefont {J.}~\bibnamefont {Bowers}},\ and\ \bibinfo
  {author} {\bibfnamefont {A.}~\bibnamefont {Mitchell}},\ }\bibfield  {title}
  {\bibinfo {title} {Lithium niobate photonics: Unlocking the electromagnetic
  spectrum},\ }\href {https://doi.org/10.1126/science.abj4396} {\bibfield
  {journal} {\bibinfo  {journal} {Science}\ }\textbf {\bibinfo {volume}
  {379}},\ \bibinfo {pages} {eabj4396} (\bibinfo {year} {2023})},\ \Eprint
  {https://arxiv.org/abs/https://www.science.org/doi/pdf/10.1126/science.abj4396}
  {https://www.science.org/doi/pdf/10.1126/science.abj4396} \BibitemShut
  {NoStop}%
\bibitem [{\citenamefont {Ruske}\ \emph {et~al.}(2003)\citenamefont {Ruske},
  \citenamefont {Zeitner}, \citenamefont {Tunnermann},\ and\ \citenamefont
  {Rasch}}]{PE2}%
  \BibitemOpen
  \bibfield  {author} {\bibinfo {author} {\bibfnamefont {J.-P.}\ \bibnamefont
  {Ruske}}, \bibinfo {author} {\bibfnamefont {B.}~\bibnamefont {Zeitner}},
  \bibinfo {author} {\bibfnamefont {A.}~\bibnamefont {Tunnermann}},\ and\
  \bibinfo {author} {\bibfnamefont {A.}~\bibnamefont {Rasch}},\ }\bibfield
  {title} {\bibinfo {title} {Photorefractive effect and high power transmission
  in linbo3 channel waveguides},\ }\href {https://doi.org/10.1049/el:20030703}
  {\bibfield  {journal} {\bibinfo  {journal} {Electronics Letters}\ }\textbf
  {\bibinfo {volume} {39}},\ \bibinfo {pages} {1048 } (\bibinfo {year}
  {2003})}\BibitemShut {NoStop}%
\bibitem [{\citenamefont {Harvey}(1988)}]{PE4}%
  \BibitemOpen
  \bibfield  {author} {\bibinfo {author} {\bibfnamefont {G.}~\bibnamefont
  {Harvey}},\ }\bibfield  {title} {\bibinfo {title} {The photorefractive effect
  in directional coupler and mach-zehnder linbo/sub 3/ optical modulators at a
  wavelength of 1.3 mu m},\ }\href {https://doi.org/10.1109/50.4075} {\bibfield
   {journal} {\bibinfo  {journal} {Journal of Lightwave Technology}\ }\textbf
  {\bibinfo {volume} {6}},\ \bibinfo {pages} {872} (\bibinfo {year}
  {1988})}\BibitemShut {NoStop}%
\bibitem [{\citenamefont {Glass}(1978)}]{PE5}%
  \BibitemOpen
  \bibfield  {author} {\bibinfo {author} {\bibfnamefont {A.~M.}\ \bibnamefont
  {Glass}},\ }\bibfield  {title} {\bibinfo {title} {{The Photorefractive
  Effect}},\ }\href {https://doi.org/10.1117/12.7972267} {\bibfield  {journal}
  {\bibinfo  {journal} {Optical Engineering}\ }\textbf {\bibinfo {volume}
  {17}},\ \bibinfo {pages} {175470} (\bibinfo {year} {1978})}\BibitemShut
  {NoStop}%
\bibitem [{\citenamefont {Kostritskii}(2009)}]{Mechanism}%
  \BibitemOpen
  \bibfield  {author} {\bibinfo {author} {\bibfnamefont {S.~M.}\ \bibnamefont
  {Kostritskii}},\ }\bibfield  {title} {\bibinfo {title} {Photorefractive
  effect in linbo3-based integrated-optical circuits at wavelengths of third
  telecom window},\ }\href {https://doi.org/10.1007/s00340-009-3501-4}
  {\bibfield  {journal} {\bibinfo  {journal} {Applied Physics B}\ }\textbf
  {\bibinfo {volume} {95}},\ \bibinfo {pages} {421} (\bibinfo {year}
  {2009})}\BibitemShut {NoStop}%
\bibitem [{\citenamefont {Buse}\ \emph {et~al.}(1998)\citenamefont {Buse},
  \citenamefont {Adibi},\ and\ \citenamefont {Psaltis}}]{PE-Positive}%
  \BibitemOpen
  \bibfield  {author} {\bibinfo {author} {\bibfnamefont {K.}~\bibnamefont
  {Buse}}, \bibinfo {author} {\bibfnamefont {A.}~\bibnamefont {Adibi}},\ and\
  \bibinfo {author} {\bibfnamefont {D.}~\bibnamefont {Psaltis}},\ }\bibfield
  {title} {\bibinfo {title} {Non-volatile holographic storage in doubly doped
  lithium niobate crystals},\ }\href {https://doi.org/10.1038/31429} {\bibfield
   {journal} {\bibinfo  {journal} {Nature}\ }\textbf {\bibinfo {volume}
  {393}},\ \bibinfo {pages} {665} (\bibinfo {year} {1998})}\BibitemShut
  {NoStop}%
\bibitem [{\citenamefont {Ye}\ \emph {et~al.}(2023)\citenamefont {Ye},
  \citenamefont {Chen}, \citenamefont {Zhang}, \citenamefont {Lu},
  \citenamefont {Wang}, \citenamefont {Huang}, \citenamefont {Wang},
  \citenamefont {He}, \citenamefont {Yin}, \citenamefont {Guo},\ and\
  \citenamefont {Han}}]{Attack-PE-1}%
  \BibitemOpen
  \bibfield  {author} {\bibinfo {author} {\bibfnamefont {P.}~\bibnamefont
  {Ye}}, \bibinfo {author} {\bibfnamefont {W.}~\bibnamefont {Chen}}, \bibinfo
  {author} {\bibfnamefont {G.-W.}\ \bibnamefont {Zhang}}, \bibinfo {author}
  {\bibfnamefont {F.-Y.}\ \bibnamefont {Lu}}, \bibinfo {author} {\bibfnamefont
  {F.-X.}\ \bibnamefont {Wang}}, \bibinfo {author} {\bibfnamefont {G.-Z.}\
  \bibnamefont {Huang}}, \bibinfo {author} {\bibfnamefont {S.}~\bibnamefont
  {Wang}}, \bibinfo {author} {\bibfnamefont {D.-Y.}\ \bibnamefont {He}},
  \bibinfo {author} {\bibfnamefont {Z.-Q.}\ \bibnamefont {Yin}}, \bibinfo
  {author} {\bibfnamefont {G.-C.}\ \bibnamefont {Guo}},\ and\ \bibinfo {author}
  {\bibfnamefont {Z.-F.}\ \bibnamefont {Han}},\ }\bibfield  {title} {\bibinfo
  {title} {Induced-photorefraction attack against quantum key distribution},\
  }\href {https://doi.org/10.1103/PhysRevApplied.19.054052} {\bibfield
  {journal} {\bibinfo  {journal} {Phys. Rev. Appl.}\ }\textbf {\bibinfo
  {volume} {19}},\ \bibinfo {pages} {054052} (\bibinfo {year}
  {2023})}\BibitemShut {NoStop}%
\bibitem [{\citenamefont {Lu}\ \emph {et~al.}(2023)\citenamefont {Lu},
  \citenamefont {Ye}, \citenamefont {Wang}, \citenamefont {Wang}, \citenamefont
  {Yin}, \citenamefont {Wang}, \citenamefont {Huang}, \citenamefont {Chen},
  \citenamefont {He}, \citenamefont {Fan-Yuan}, \citenamefont {Guo},\ and\
  \citenamefont {Han}}]{Attack-PE-2}%
  \BibitemOpen
  \bibfield  {author} {\bibinfo {author} {\bibfnamefont {F.-Y.}\ \bibnamefont
  {Lu}}, \bibinfo {author} {\bibfnamefont {P.}~\bibnamefont {Ye}}, \bibinfo
  {author} {\bibfnamefont {Z.-H.}\ \bibnamefont {Wang}}, \bibinfo {author}
  {\bibfnamefont {S.}~\bibnamefont {Wang}}, \bibinfo {author} {\bibfnamefont
  {Z.-Q.}\ \bibnamefont {Yin}}, \bibinfo {author} {\bibfnamefont
  {R.}~\bibnamefont {Wang}}, \bibinfo {author} {\bibfnamefont {X.-J.}\
  \bibnamefont {Huang}}, \bibinfo {author} {\bibfnamefont {W.}~\bibnamefont
  {Chen}}, \bibinfo {author} {\bibfnamefont {D.-Y.}\ \bibnamefont {He}},
  \bibinfo {author} {\bibfnamefont {G.-J.}\ \bibnamefont {Fan-Yuan}}, \bibinfo
  {author} {\bibfnamefont {G.-C.}\ \bibnamefont {Guo}},\ and\ \bibinfo {author}
  {\bibfnamefont {Z.-F.}\ \bibnamefont {Han}},\ }\bibfield  {title} {\bibinfo
  {title} {Hacking measurement-device-independent quantum key distribution},\
  }\href {https://doi.org/10.1364/OPTICA.485389} {\bibfield  {journal}
  {\bibinfo  {journal} {Optica}\ }\textbf {\bibinfo {volume} {10}},\ \bibinfo
  {pages} {520} (\bibinfo {year} {2023})}\BibitemShut {NoStop}%
\bibitem [{\citenamefont {Hall}\ \emph {et~al.}(1985)\citenamefont {Hall},
  \citenamefont {Jaura}, \citenamefont {Connors},\ and\ \citenamefont
  {Foote}}]{PE3}%
  \BibitemOpen
  \bibfield  {author} {\bibinfo {author} {\bibfnamefont {T.}~\bibnamefont
  {Hall}}, \bibinfo {author} {\bibfnamefont {R.}~\bibnamefont {Jaura}},
  \bibinfo {author} {\bibfnamefont {L.}~\bibnamefont {Connors}},\ and\ \bibinfo
  {author} {\bibfnamefont {P.}~\bibnamefont {Foote}},\ }\bibfield  {title}
  {\bibinfo {title} {The photorefractive effect—a review},\ }\href
  {https://doi.org/https://doi.org/10.1016/0079-6727(85)90001-1} {\bibfield
  {journal} {\bibinfo  {journal} {Progress in Quantum Electronics}\ }\textbf
  {\bibinfo {volume} {10}},\ \bibinfo {pages} {77} (\bibinfo {year}
  {1985})}\BibitemShut {NoStop}%
\bibitem [{\citenamefont {Becker}\ and\ \citenamefont
  {Williamson}(1985)}]{PE1}%
  \BibitemOpen
  \bibfield  {author} {\bibinfo {author} {\bibfnamefont {R.~A.}\ \bibnamefont
  {Becker}}\ and\ \bibinfo {author} {\bibfnamefont {R.~C.}\ \bibnamefont
  {Williamson}},\ }\bibfield  {title} {\bibinfo {title} {{Photorefractive
  effects in LiNbO3 channel waveguides: Model and experimental verification}},\
  }\href {https://doi.org/10.1063/1.96365} {\bibfield  {journal} {\bibinfo
  {journal} {Applied Physics Letters}\ }\textbf {\bibinfo {volume} {47}},\
  \bibinfo {pages} {1024} (\bibinfo {year} {1985})},\ \Eprint
  {https://arxiv.org/abs/https://pubs.aip.org/aip/apl/article-pdf/47/10/1024/18455526/1024\_1\_online.pdf}
  {https://pubs.aip.org/aip/apl/article-pdf/47/10/1024/18455526/1024\_1\_online.pdf}
  \BibitemShut {NoStop}%
\bibitem [{\citenamefont {N.~V.~Kukhtarev}\ and\ \citenamefont
  {Vinetskii}(1978)}]{Kukhtarev}%
  \BibitemOpen
  \bibfield  {author} {\bibinfo {author} {\bibfnamefont {S.~G. O. M. S.~S.}\
  \bibnamefont {N.~V.~Kukhtarev}, \bibfnamefont {V.~B.~Markov}}\ and\ \bibinfo
  {author} {\bibfnamefont {V.~L.}\ \bibnamefont {Vinetskii}},\ }\bibfield
  {title} {\bibinfo {title} {Holographic storage in electrooptic crystals. i.
  steady state},\ }\href {https://doi.org/10.1080/00150197908239450} {\bibfield
   {journal} {\bibinfo  {journal} {Ferroelectrics}\ }\textbf {\bibinfo {volume}
  {22}},\ \bibinfo {pages} {949} (\bibinfo {year} {1978})},\ \Eprint
  {https://arxiv.org/abs/https://doi.org/10.1080/00150197908239450}
  {https://doi.org/10.1080/00150197908239450} \BibitemShut {NoStop}%
\bibitem [{\citenamefont {Pan}\ \emph {et~al.}(2022)\citenamefont {Pan},
  \citenamefont {Wang}, \citenamefont {Shao}, \citenamefont {Pi}, \citenamefont
  {Li}, \citenamefont {Liu}, \citenamefont {Huang},\ and\ \citenamefont
  {Xu}}]{Implementation1}%
  \BibitemOpen
  \bibfield  {author} {\bibinfo {author} {\bibfnamefont {Y.}~\bibnamefont
  {Pan}}, \bibinfo {author} {\bibfnamefont {H.}~\bibnamefont {Wang}}, \bibinfo
  {author} {\bibfnamefont {Y.}~\bibnamefont {Shao}}, \bibinfo {author}
  {\bibfnamefont {Y.}~\bibnamefont {Pi}}, \bibinfo {author} {\bibfnamefont
  {Y.}~\bibnamefont {Li}}, \bibinfo {author} {\bibfnamefont {B.}~\bibnamefont
  {Liu}}, \bibinfo {author} {\bibfnamefont {W.}~\bibnamefont {Huang}},\ and\
  \bibinfo {author} {\bibfnamefont {B.}~\bibnamefont {Xu}},\ }\bibfield
  {title} {\bibinfo {title} {Experimental demonstration of high-rate
  discrete-modulated continuous-variable quantum key distribution system},\
  }\href {https://doi.org/10.1364/OL.456978} {\bibfield  {journal} {\bibinfo
  {journal} {Opt. Lett.}\ }\textbf {\bibinfo {volume} {47}},\ \bibinfo {pages}
  {3307} (\bibinfo {year} {2022})}\BibitemShut {NoStop}%
\bibitem [{\citenamefont {Ji}\ \emph {et~al.}(2024)\citenamefont {Ji},
  \citenamefont {Huang}, \citenamefont {Wang}, \citenamefont {Jiang},\ and\
  \citenamefont {Zeng}}]{Implementation2}%
  \BibitemOpen
  \bibfield  {author} {\bibinfo {author} {\bibfnamefont {F.}~\bibnamefont
  {Ji}}, \bibinfo {author} {\bibfnamefont {P.}~\bibnamefont {Huang}}, \bibinfo
  {author} {\bibfnamefont {T.}~\bibnamefont {Wang}}, \bibinfo {author}
  {\bibfnamefont {X.}~\bibnamefont {Jiang}},\ and\ \bibinfo {author}
  {\bibfnamefont {G.}~\bibnamefont {Zeng}},\ }\bibfield  {title} {\bibinfo
  {title} {Gbps key rate passive-state-preparation continuous-variable quantum
  key distribution within an access-network area},\ }\href
  {https://doi.org/10.1364/PRJ.519909} {\bibfield  {journal} {\bibinfo
  {journal} {Photon. Res.}\ }\textbf {\bibinfo {volume} {12}},\ \bibinfo
  {pages} {1485} (\bibinfo {year} {2024})}\BibitemShut {NoStop}%
\bibitem [{\citenamefont {Huang}\ \emph {et~al.}(2016)\citenamefont {Huang},
  \citenamefont {Huang}, \citenamefont {Lin},\ and\ \citenamefont
  {Zeng}}]{Huang2016}%
  \BibitemOpen
  \bibfield  {author} {\bibinfo {author} {\bibfnamefont {D.}~\bibnamefont
  {Huang}}, \bibinfo {author} {\bibfnamefont {P.}~\bibnamefont {Huang}},
  \bibinfo {author} {\bibfnamefont {D.}~\bibnamefont {Lin}},\ and\ \bibinfo
  {author} {\bibfnamefont {G.}~\bibnamefont {Zeng}},\ }\bibfield  {title}
  {\bibinfo {title} {Long-distance continuous-variable quantum key distribution
  by controlling excess noise},\ }\href {https://doi.org/10.1038/srep19201}
  {\bibfield  {journal} {\bibinfo  {journal} {Scientific Reports}\ }\textbf
  {\bibinfo {volume} {6}},\ \bibinfo {pages} {19201} (\bibinfo {year}
  {2016})}\BibitemShut {NoStop}%
\bibitem [{\citenamefont {Zhang}\ \emph {et~al.}(2020)\citenamefont {Zhang},
  \citenamefont {Chen}, \citenamefont {Pirandola}, \citenamefont {Wang},
  \citenamefont {Zhou}, \citenamefont {Chu}, \citenamefont {Zhao},
  \citenamefont {Xu}, \citenamefont {Yu},\ and\ \citenamefont
  {Guo}}]{PhysRevLett.125.010502}%
  \BibitemOpen
  \bibfield  {author} {\bibinfo {author} {\bibfnamefont {Y.}~\bibnamefont
  {Zhang}}, \bibinfo {author} {\bibfnamefont {Z.}~\bibnamefont {Chen}},
  \bibinfo {author} {\bibfnamefont {S.}~\bibnamefont {Pirandola}}, \bibinfo
  {author} {\bibfnamefont {X.}~\bibnamefont {Wang}}, \bibinfo {author}
  {\bibfnamefont {C.}~\bibnamefont {Zhou}}, \bibinfo {author} {\bibfnamefont
  {B.}~\bibnamefont {Chu}}, \bibinfo {author} {\bibfnamefont {Y.}~\bibnamefont
  {Zhao}}, \bibinfo {author} {\bibfnamefont {B.}~\bibnamefont {Xu}}, \bibinfo
  {author} {\bibfnamefont {S.}~\bibnamefont {Yu}},\ and\ \bibinfo {author}
  {\bibfnamefont {H.}~\bibnamefont {Guo}},\ }\bibfield  {title} {\bibinfo
  {title} {Long-distance continuous-variable quantum key distribution over
  202.81 km of fiber},\ }\href {https://doi.org/10.1103/PhysRevLett.125.010502}
  {\bibfield  {journal} {\bibinfo  {journal} {Phys. Rev. Lett.}\ }\textbf
  {\bibinfo {volume} {125}},\ \bibinfo {pages} {010502} (\bibinfo {year}
  {2020})}\BibitemShut {NoStop}%
\bibitem [{\citenamefont {Jouguet}\ \emph
  {et~al.}(2013{\natexlab{b}})\citenamefont {Jouguet}, \citenamefont
  {Kunz-Jacques}, \citenamefont {Leverrier}, \citenamefont {Grangier},\ and\
  \citenamefont {Diamanti}}]{Jouguet2013}%
  \BibitemOpen
  \bibfield  {author} {\bibinfo {author} {\bibfnamefont {P.}~\bibnamefont
  {Jouguet}}, \bibinfo {author} {\bibfnamefont {S.}~\bibnamefont
  {Kunz-Jacques}}, \bibinfo {author} {\bibfnamefont {A.}~\bibnamefont
  {Leverrier}}, \bibinfo {author} {\bibfnamefont {P.}~\bibnamefont
  {Grangier}},\ and\ \bibinfo {author} {\bibfnamefont {E.}~\bibnamefont
  {Diamanti}},\ }\bibfield  {title} {\bibinfo {title} {Experimental
  demonstration of long-distance continuous-variable quantum key
  distribution},\ }\href {https://doi.org/10.1038/nphoton.2013.63} {\bibfield
  {journal} {\bibinfo  {journal} {Nature Photonics}\ }\textbf {\bibinfo
  {volume} {7}},\ \bibinfo {pages} {378} (\bibinfo {year}
  {2013}{\natexlab{b}})}\BibitemShut {NoStop}%
\bibitem [{\citenamefont {Fossier}\ \emph {et~al.}(2009)\citenamefont
  {Fossier}, \citenamefont {Diamanti}, \citenamefont {Debuisschert},
  \citenamefont {Tualle-Brouri},\ and\ \citenamefont {Grangier}}]{amplifier}%
  \BibitemOpen
  \bibfield  {author} {\bibinfo {author} {\bibfnamefont {S.}~\bibnamefont
  {Fossier}}, \bibinfo {author} {\bibfnamefont {E.}~\bibnamefont {Diamanti}},
  \bibinfo {author} {\bibfnamefont {T.}~\bibnamefont {Debuisschert}}, \bibinfo
  {author} {\bibfnamefont {R.}~\bibnamefont {Tualle-Brouri}},\ and\ \bibinfo
  {author} {\bibfnamefont {P.}~\bibnamefont {Grangier}},\ }\bibfield  {title}
  {\bibinfo {title} {Improvement of continuous-variable quantum key
  distribution systems by using optical preamplifiers},\ }\href
  {https://doi.org/10.1088/0953-4075/42/11/114014} {\bibfield  {journal}
  {\bibinfo  {journal} {Journal of Physics B: Atomic, Molecular and Optical
  Physics}\ }\textbf {\bibinfo {volume} {42}},\ \bibinfo {pages} {114014}
  (\bibinfo {year} {2009})}\BibitemShut {NoStop}%
\bibitem [{\citenamefont {Leverrier}\ \emph {et~al.}(2010)\citenamefont
  {Leverrier}, \citenamefont {Grosshans},\ and\ \citenamefont
  {Grangier}}]{Finitesize-1}%
  \BibitemOpen
  \bibfield  {author} {\bibinfo {author} {\bibfnamefont {A.}~\bibnamefont
  {Leverrier}}, \bibinfo {author} {\bibfnamefont {F.}~\bibnamefont
  {Grosshans}},\ and\ \bibinfo {author} {\bibfnamefont {P.}~\bibnamefont
  {Grangier}},\ }\bibfield  {title} {\bibinfo {title} {Finite-size analysis of
  a continuous-variable quantum key distribution},\ }\href
  {https://doi.org/10.1103/PhysRevA.81.062343} {\bibfield  {journal} {\bibinfo
  {journal} {Phys. Rev. A}\ }\textbf {\bibinfo {volume} {81}},\ \bibinfo
  {pages} {062343} (\bibinfo {year} {2010})}\BibitemShut {NoStop}%
\bibitem [{\citenamefont {Leverrier}\ \emph {et~al.}(2013)\citenamefont
  {Leverrier}, \citenamefont {Garc\'{\i}a-Patr\'on}, \citenamefont {Renner},\
  and\ \citenamefont {Cerf}}]{Finitesize-2}%
  \BibitemOpen
  \bibfield  {author} {\bibinfo {author} {\bibfnamefont {A.}~\bibnamefont
  {Leverrier}}, \bibinfo {author} {\bibfnamefont {R.}~\bibnamefont
  {Garc\'{\i}a-Patr\'on}}, \bibinfo {author} {\bibfnamefont {R.}~\bibnamefont
  {Renner}},\ and\ \bibinfo {author} {\bibfnamefont {N.~J.}\ \bibnamefont
  {Cerf}},\ }\bibfield  {title} {\bibinfo {title} {Security of
  continuous-variable quantum key distribution against general attacks},\
  }\href {https://doi.org/10.1103/PhysRevLett.110.030502} {\bibfield  {journal}
  {\bibinfo  {journal} {Phys. Rev. Lett.}\ }\textbf {\bibinfo {volume} {110}},\
  \bibinfo {pages} {030502} (\bibinfo {year} {2013})}\BibitemShut {NoStop}%
\bibitem [{\citenamefont {Fu}\ \emph {et~al.}(2013{\natexlab{b}})\citenamefont
  {Fu}, \citenamefont {Zhang}, \citenamefont {Hraimel}, \citenamefont {Liu},\
  and\ \citenamefont {Shen}}]{biaslocking}%
  \BibitemOpen
  \bibfield  {author} {\bibinfo {author} {\bibfnamefont {Y.}~\bibnamefont
  {Fu}}, \bibinfo {author} {\bibfnamefont {X.}~\bibnamefont {Zhang}}, \bibinfo
  {author} {\bibfnamefont {B.}~\bibnamefont {Hraimel}}, \bibinfo {author}
  {\bibfnamefont {T.}~\bibnamefont {Liu}},\ and\ \bibinfo {author}
  {\bibfnamefont {D.}~\bibnamefont {Shen}},\ }\bibfield  {title} {\bibinfo
  {title} {Mach-zehnder: A review of bias control techniques for mach-zehnder
  modulators in photonic analog links},\ }\href
  {https://doi.org/10.1109/MMM.2013.2280332} {\bibfield  {journal} {\bibinfo
  {journal} {IEEE Microwave Magazine}\ }\textbf {\bibinfo {volume} {14}},\
  \bibinfo {pages} {102} (\bibinfo {year} {2013}{\natexlab{b}})}\BibitemShut
  {NoStop}%
\bibitem [{\citenamefont {{K{\"o}sters}}\ \emph {et~al.}(2009)\citenamefont
  {{K{\"o}sters}}, \citenamefont {{Sturman}}, \citenamefont {{Werheit}},
  \citenamefont {{Haertle}},\ and\ \citenamefont {{Buse}}}]{cleaning}%
  \BibitemOpen
  \bibfield  {author} {\bibinfo {author} {\bibfnamefont {M.}~\bibnamefont
  {{K{\"o}sters}}}, \bibinfo {author} {\bibfnamefont {B.}~\bibnamefont
  {{Sturman}}}, \bibinfo {author} {\bibfnamefont {P.}~\bibnamefont
  {{Werheit}}}, \bibinfo {author} {\bibfnamefont {D.}~\bibnamefont
  {{Haertle}}},\ and\ \bibinfo {author} {\bibfnamefont {K.}~\bibnamefont
  {{Buse}}},\ }\bibfield  {title} {\bibinfo {title} {{Optical cleaning of
  congruent lithium niobate crystals}},\ }\href
  {https://doi.org/10.1038/nphoton.2009.142} {\bibfield  {journal} {\bibinfo
  {journal} {Nature Photonics}\ }\textbf {\bibinfo {volume} {3}},\ \bibinfo
  {pages} {510} (\bibinfo {year} {2009})}\BibitemShut {NoStop}%
\bibitem [{\citenamefont {Kong}\ \emph {et~al.}(2019)\citenamefont {Kong},
  \citenamefont {Bo}, \citenamefont {Wang}, \citenamefont {Zheng},
  \citenamefont {Liu}, \citenamefont {Zhang}, \citenamefont {Rupp},\ and\
  \citenamefont {Xu}}]{Doping3}%
  \BibitemOpen
  \bibfield  {author} {\bibinfo {author} {\bibfnamefont {Y.}~\bibnamefont
  {Kong}}, \bibinfo {author} {\bibfnamefont {F.}~\bibnamefont {Bo}}, \bibinfo
  {author} {\bibfnamefont {W.}~\bibnamefont {Wang}}, \bibinfo {author}
  {\bibfnamefont {D.}~\bibnamefont {Zheng}}, \bibinfo {author} {\bibfnamefont
  {H.}~\bibnamefont {Liu}}, \bibinfo {author} {\bibfnamefont {G.}~\bibnamefont
  {Zhang}}, \bibinfo {author} {\bibfnamefont {R.}~\bibnamefont {Rupp}},\ and\
  \bibinfo {author} {\bibfnamefont {J.}~\bibnamefont {Xu}},\ }\bibfield
  {title} {\bibinfo {title} {Recent progress in lithium niobate: Optical
  damage, defect simulation, and on-chip devices},\ }\href
  {https://doi.org/10.1002/adma.201806452} {\bibfield  {journal} {\bibinfo
  {journal} {Advanced Materials}\ }\textbf {\bibinfo {volume} {32}},\ \bibinfo
  {pages} {1806452} (\bibinfo {year} {2019})}\BibitemShut {NoStop}%
\bibitem [{\citenamefont {Eichmann}(2016)}]{Doping4}%
  \BibitemOpen
  \bibfield  {author} {\bibinfo {author} {\bibfnamefont {D.}~\bibnamefont
  {Eichmann}},\ }\bibfield  {title} {\bibinfo {title} {Lithium niobate defects
  photorefraction and ferroelectric switching}\ }(\bibinfo {year}
  {2016})\BibitemShut {NoStop}%
\bibitem [{\citenamefont {Kong}\ \emph {et~al.}(2012)\citenamefont {Kong},
  \citenamefont {Liu},\ and\ \citenamefont {Xu}}]{Doping1}%
  \BibitemOpen
  \bibfield  {author} {\bibinfo {author} {\bibfnamefont {Y.}~\bibnamefont
  {Kong}}, \bibinfo {author} {\bibfnamefont {S.}~\bibnamefont {Liu}},\ and\
  \bibinfo {author} {\bibfnamefont {J.}~\bibnamefont {Xu}},\ }\bibfield
  {title} {\bibinfo {title} {Recent advances in the photorefraction of doped
  lithium niobate crystals},\ }\href {https://doi.org/10.3390/ma5101954}
  {\bibfield  {journal} {\bibinfo  {journal} {Materials}\ }\textbf {\bibinfo
  {volume} {5}},\ \bibinfo {pages} {1954} (\bibinfo {year} {2012})}\BibitemShut
  {NoStop}%
\bibitem [{\citenamefont {Liu}\ \emph {et~al.}(2006)\citenamefont {Liu},
  \citenamefont {Xie}, \citenamefont {Kong}, \citenamefont {Yan}, \citenamefont
  {Li}, \citenamefont {Shi}, \citenamefont {Xu},\ and\ \citenamefont
  {Zhang}}]{Doping2}%
  \BibitemOpen
  \bibfield  {author} {\bibinfo {author} {\bibfnamefont {H.}~\bibnamefont
  {Liu}}, \bibinfo {author} {\bibfnamefont {X.}~\bibnamefont {Xie}}, \bibinfo
  {author} {\bibfnamefont {Y.}~\bibnamefont {Kong}}, \bibinfo {author}
  {\bibfnamefont {W.}~\bibnamefont {Yan}}, \bibinfo {author} {\bibfnamefont
  {X.}~\bibnamefont {Li}}, \bibinfo {author} {\bibfnamefont {L.}~\bibnamefont
  {Shi}}, \bibinfo {author} {\bibfnamefont {J.}~\bibnamefont {Xu}},\ and\
  \bibinfo {author} {\bibfnamefont {G.}~\bibnamefont {Zhang}},\ }\bibfield
  {title} {\bibinfo {title} {Photorefractive properties of near-stoichiometric
  lithium niobate crystals doped with iron},\ }\href
  {https://doi.org/https://doi.org/10.1016/j.optmat.2004.12.018} {\bibfield
  {journal} {\bibinfo  {journal} {Optical Materials}\ }\textbf {\bibinfo
  {volume} {28}},\ \bibinfo {pages} {212} (\bibinfo {year} {2006})}\BibitemShut
  {NoStop}%
\bibitem [{\citenamefont {Grosshans}\ and\ \citenamefont
  {Grangier}(2002{\natexlab{b}})}]{GG02}%
  \BibitemOpen
  \bibfield  {author} {\bibinfo {author} {\bibfnamefont {F.}~\bibnamefont
  {Grosshans}}\ and\ \bibinfo {author} {\bibfnamefont {P.}~\bibnamefont
  {Grangier}},\ }\bibfield  {title} {\bibinfo {title} {Continuous variable
  quantum cryptography using coherent states},\ }\href
  {https://doi.org/10.1103/PhysRevLett.88.057902} {\bibfield  {journal}
  {\bibinfo  {journal} {Phys. Rev. Lett.}\ }\textbf {\bibinfo {volume} {88}},\
  \bibinfo {pages} {057902} (\bibinfo {year} {2002}{\natexlab{b}})}\BibitemShut
  {NoStop}%
\bibitem [{\citenamefont {Renner}\ and\ \citenamefont {Cirac}(2009)}]{11}%
  \BibitemOpen
  \bibfield  {author} {\bibinfo {author} {\bibfnamefont {R.}~\bibnamefont
  {Renner}}\ and\ \bibinfo {author} {\bibfnamefont {J.~I.}\ \bibnamefont
  {Cirac}},\ }\bibfield  {title} {\bibinfo {title} {de finetti representation
  theorem for infinite-dimensional quantum systems and applications to quantum
  cryptography},\ }\href {https://doi.org/10.1103/PhysRevLett.102.110504}
  {\bibfield  {journal} {\bibinfo  {journal} {Phys. Rev. Lett.}\ }\textbf
  {\bibinfo {volume} {102}},\ \bibinfo {pages} {110504} (\bibinfo {year}
  {2009})}\BibitemShut {NoStop}%
\bibitem [{\citenamefont {Leverrier}\ and\ \citenamefont
  {Grangier}(2009)}]{12}%
  \BibitemOpen
  \bibfield  {author} {\bibinfo {author} {\bibfnamefont {A.}~\bibnamefont
  {Leverrier}}\ and\ \bibinfo {author} {\bibfnamefont {P.}~\bibnamefont
  {Grangier}},\ }\bibfield  {title} {\bibinfo {title} {Unconditional security
  proof of long-distance continuous-variable quantum key distribution with
  discrete modulation},\ }\href
  {https://doi.org/10.1103/PhysRevLett.102.180504} {\bibfield  {journal}
  {\bibinfo  {journal} {Phys. Rev. Lett.}\ }\textbf {\bibinfo {volume} {102}},\
  \bibinfo {pages} {180504} (\bibinfo {year} {2009})}\BibitemShut {NoStop}%
\bibitem [{\citenamefont {Lodewyck}\ \emph {et~al.}(2005)\citenamefont
  {Lodewyck}, \citenamefont {Debuisschert}, \citenamefont {Tualle-Brouri},\
  and\ \citenamefont {Grangier}}]{14}%
  \BibitemOpen
  \bibfield  {author} {\bibinfo {author} {\bibfnamefont {J.}~\bibnamefont
  {Lodewyck}}, \bibinfo {author} {\bibfnamefont {T.}~\bibnamefont
  {Debuisschert}}, \bibinfo {author} {\bibfnamefont {R.}~\bibnamefont
  {Tualle-Brouri}},\ and\ \bibinfo {author} {\bibfnamefont {P.}~\bibnamefont
  {Grangier}},\ }\bibfield  {title} {\bibinfo {title} {Controlling excess noise
  in fiber-optics continuous-variable quantum key distribution},\ }\href
  {https://doi.org/10.1103/PhysRevA.72.050303} {\bibfield  {journal} {\bibinfo
  {journal} {Phys. Rev. A}\ }\textbf {\bibinfo {volume} {72}},\ \bibinfo
  {pages} {050303} (\bibinfo {year} {2005})}\BibitemShut {NoStop}%
\bibitem [{\citenamefont {Jouguet}\ \emph
  {et~al.}(2013{\natexlab{c}})\citenamefont {Jouguet}, \citenamefont
  {Kunz-Jacques}, \citenamefont {Leverrier}, \citenamefont {Grangier},\ and\
  \citenamefont {Diamanti}}]{15}%
  \BibitemOpen
  \bibfield  {author} {\bibinfo {author} {\bibfnamefont {P.}~\bibnamefont
  {Jouguet}}, \bibinfo {author} {\bibfnamefont {S.}~\bibnamefont
  {Kunz-Jacques}}, \bibinfo {author} {\bibfnamefont {A.}~\bibnamefont
  {Leverrier}}, \bibinfo {author} {\bibfnamefont {P.}~\bibnamefont
  {Grangier}},\ and\ \bibinfo {author} {\bibfnamefont {E.}~\bibnamefont
  {Diamanti}},\ }\bibfield  {title} {\bibinfo {title} {Experimental
  demonstration of long-distance continuous-variable quantum key
  distribution},\ }\href {https://doi.org/10.1038/nphoton.2013.63} {\bibfield
  {journal} {\bibinfo  {journal} {Nature Photonics}\ }\textbf {\bibinfo
  {volume} {7}},\ \bibinfo {pages} {378–381} (\bibinfo {year}
  {2013}{\natexlab{c}})}\BibitemShut {NoStop}%
\bibitem [{\citenamefont {Sudjana}\ \emph {et~al.}(2007)\citenamefont
  {Sudjana}, \citenamefont {Magnin}, \citenamefont {Garc{\'i}a-Patr{\'o}n},\
  and\ \citenamefont {Cerf}}]{GMCS-Het-Re-Ind-3}%
  \BibitemOpen
  \bibfield  {author} {\bibinfo {author} {\bibfnamefont {J.}~\bibnamefont
  {Sudjana}}, \bibinfo {author} {\bibfnamefont {L.}~\bibnamefont {Magnin}},
  \bibinfo {author} {\bibfnamefont {R.}~\bibnamefont {Garc{\'i}a-Patr{\'o}n}},\
  and\ \bibinfo {author} {\bibfnamefont {N.~J.}\ \bibnamefont {Cerf}},\
  }\bibfield  {title} {\bibinfo {title} {Tight bounds on the eavesdropping of a
  continuous-variable quantum cryptographic protocol with no basis switching},\
  }\href {https://api.semanticscholar.org/CorpusID:55931548} {\bibfield
  {journal} {\bibinfo  {journal} {Physical Review A}\ }\textbf {\bibinfo
  {volume} {76}},\ \bibinfo {pages} {052301} (\bibinfo {year}
  {2007})}\BibitemShut {NoStop}%
\bibitem [{\citenamefont {Leverrier}(2017{\natexlab{b}})}]{Finitesize-3}%
  \BibitemOpen
  \bibfield  {author} {\bibinfo {author} {\bibfnamefont {A.}~\bibnamefont
  {Leverrier}},\ }\bibfield  {title} {\bibinfo {title} {Security of
  continuous-variable quantum key distribution via a gaussian de finetti
  reduction},\ }\bibfield  {journal} {\bibinfo  {journal} {Physical Review
  Letters}\ }\textbf {\bibinfo {volume} {118}},\ \href
  {https://doi.org/10.1103/physrevlett.118.200501}
  {10.1103/physrevlett.118.200501} (\bibinfo {year}
  {2017}{\natexlab{b}})\BibitemShut {NoStop}%
\end{thebibliography}%
\end{document}